\documentclass[a4paper,11pt]{article}
\usepackage{jheppub} 
\usepackage{lineno}
\usepackage{soul}
\usepackage{xcolor}

\title{\boldmath Impact of unitarity violation on sensitivity of the leptonic CP phase at Hyper-Kamiokande and DUNE}

\author{Ana Maria Garcia Trzeciak,$^{a}$ Hiroshi Nunokawa,$^{a}$ and Alexander A. Quiroga$^{b}$}
\affiliation{$^{a}$Departamento de Física, Pontifícia Universidade Católica do Rio de Janeiro,\\
C.P. 38097, 22451-900, Rio de Janeiro, Brazil}
\affiliation{$^{b}$ILACVN, Universidade Federal da Integração Latino-Americana,\\
85870-650, Foz do Iguaçu, Brazil}

\emailAdd{anamgtrzeciak@aluno.puc-rio.br, nunokawa@puc-rio.br, alexander.quiroga@unila.edu.br}

\abstract{We study the impact of unitarity violation on the sensitivity of the leptonic CP phase, $\delta_{CP}$, considering the next generation of long-baseline neutrino experiments, Hyper-Kamiokande and DUNE. By simulating near and far detectors and assuming different scenarios for non-unitarity, we verify how it can affect the sensitivity to measure the $\delta_{ CP}$ violating phase. We also probe the capability of these experiments to constrain the non-unitarity parameters and how their capability could be improved if the impact of non-unitarity at both near and the far detectors were properly taken into account. We find that the Hyper-Kamiokande experiment is robust in the presence of non-unitarity mixing, achieving a sensitivity above $5\sigma$ for our all considered cases. On the other hand, DUNE suffers somewhat more impact due to unitarity violation, reaching a sensitivity below 5$\sigma$ for some values of $\delta_{CP}$. However, depending on the scenario adopted for non-unitarity, DUNE demonstrates robustness in the sensitivity to $\delta_{ CP}$ phase.}

\begin{document}
\maketitle
\flushbottom

\section{Introduction}
\label{sec:intro}

The discovery that neutrinos oscillate has given us a new understanding of the Standard Model (SM) of elementary particles: neutrinos are mixed and therefore have masses. This is a clear evidence of physics Beyond the Standard Model because SM does not provide a mechanism for such masses to exist. The neutrino mixing is described by the relation $\vert \nu_{\alpha}\rangle = \sum_{k} U^{*}_{\alpha k}\vert \nu_{k}\rangle$, where $\vert \nu_{\alpha}\rangle$ ($\alpha = e, \mu, \tau$) are three neutrino flavor eigenstates that participate in standard weak interactions, and each of them is a linear combination of mass eigenstates, $\vert \nu_{k}\rangle$ ($k = 1, 2, 3$), weighted by a mixing matrix $U$. This leptonic mixing matrix is generally assumed to be a 3$\times$3 unitary matrix called the Pontecorvo-Maki-Nakagawa-Sakata (PMNS) matrix and is commonly parametrized by three mixing angles $\theta_{12}$, $\theta_{13}$ and $\theta_{23}$, a Dirac phase $\delta_{CP}$ and two Majorana phases. In principle, Majorana phases do not participate in neutrino oscillations, therefore, we will exclude them from our discussion. In addition, to fully describe the neutrino oscillation, it is necessary to consider other parameters called mass squared differences, defined as $\Delta m^{2}_{ij} = m^{2}_{i} - m^{2}_{j}$ ($i, j$ = 1, 2, 3) where $m_{i}$ and $m_{j}$ are the neutrino masses and only two of $\Delta m^{2}_{ij}$ are independent. These mixing angles, CP phase and masses splinting are usually called neutrino oscillation parameters.

Neutrino experiments over the years have been measuring such oscillation parameters \cite{ParticleDataGroup:2024}, however, some of them still remain unknown. This is the case for the $\theta_{23}$ octant, the value of $\delta_{CP}$, the neutrino mass ordering (normal or inverted ordering) and the nature of neutrinos, if they are Dirac or Majorana. With new experiments under preparation, such as JUNO \cite{JUNO2016, JUNO2022}, Hyper-Kamiokande (Hyper-K or HK) \cite{T2HK} and DUNE \cite{DUNE-DESING}, we are entering in an era of precision, where the ability to measure the oscillation neutrino parameters is going to be enhanced to unprecedented accuracy. Therefore, we have a good chance to test another crucial question in neutrino physics: is the neutrino mixing matrix unitary? In other words, are there really only three neutrino flavors? If the neutrino mixing matrix is really non-unitary, this would have profound implications for our understanding of particle physics and physics beyond the Standard Model.

Models that aim to explain why the masses of neutrinos are so small compared to other fermions in the standard model allow the mixing matrix in the light neutrino sector to be non-unitary. This is the case of the seesaw mechanism \cite{MINKOWSKI1977,Yanagida:1979, gellmann1979, Mohapatra, Schechter, Glashow:1979nm}. The seesaw mechanism is particularly interesting and finds its most common realization within the simplest $\mathrm{SU}(2)_{L} \otimes \mathrm{U}(1)_{Y}$ gauge structure \cite{Schechter}. These models can also lead to new phenomenology, lepton number violation, and new particles which may be observable at the collider experiments \cite{Forero_2011, Schechter, Ibarra_2004, Hirsch_2009}. In high-scale implementation of the seesaw mechanism, neutrinos get mass due to the exchange of heavy singlet fermion mediators, with mass $M \gg$ $\mathcal{O}$(TeV). However, there is no direct evidence for a large or extremely large seesaw scale, on the other hand, models based on low-scale seesaw \cite{Mohapatra_1986,GONZALEZGARCIA_1989,Akhmedov_1996,Malinsky_2005} introduce neutral heavy leptons, with mass of the order of GeV/TeV, to generate small neutrino masses. These particles are known as sterile neutrinos, and evidence of them could be found in large detectors such as ATLAS and CMS at LHC \cite{DITTMAR_1990, GonzalezGarcia1990, Aguilar_2012,Drewes2020,CMSColab_2018,Cottin2023}. Moreover, some anomalies in the eV energy scale point toward the existence of light sterile neutrinos with mass of order $\sim$ eV: The LSND and the MiniBooNE experiments report an excess of unexplained events in the low energy region for $\bar{\nu}_{\mu} (\nu_{\mu}) \rightarrow \bar{\nu}_{e}(\nu_{e}$) \cite{LSND, MiniBoone_2007, MiniBoone_2010, Miniboone_2012, Minibone_2013}; the gallium experiments, GALLEX and SAGE, observed a lower-than-expected number of neutrino events from an intense radioactive source \cite{gallexColab_1995,gallexIVColab_1999,SAGE_1996,SAGEColab_1999,SAGE_2009,gallex_1999,BNO_LNGS_2006,Giunti_2011, Kostensalo_2019,kostensalo_2020} (the gallium anomaly); neutrino fluxes from nuclear reactors ($L \lesssim$ 100 m) are in tension with prediction \cite{Acero_2008,MullerTH_2011,huberP_2011}, (the reactor anomaly). Furthermore, the mixture between sterile neutrinos and three active neutrinos can cause deviations from unitary in the PMNS mixing matrix, which may leave signatures in the oscillation probability, and long-baseline experiments can be used to measure this deviation \cite{ESCRIHUELA,Antusch_2006}. A plethora of articles regarding the non-unitary neutrino mixing has been performed in the literature \cite{HIROSHI,Fong_2019,Miranda_2016,Li:2015oal,ESSnuSB_1,Antusch_2006,qian_2013,Escrihuela_2017, Dutta_2016, DUTTA_2017, Ge_Pasquini_2017,Soumya_2018,Giunti_2021,PARKE,Agarwalla:2021,Blennow_Pila:2016,Rahaman_2022,Ellis_Kelly:2020,Argawalla_2023,Moreno_2024,Miranda_2021,Hu2021Global,Fernandez-Martinez:2007,Antusch_2009,Antusch2014Nonunitarity,Verma2018,Meloni_2010, kozynets_2024,Gariazzo_2022, Kaur_2021,WANG_2022,MALINSKY_2009}. The impact of unitarity violation on the standard oscillation parameters can be found in refs. \cite{Escrihuela_2017, Dutta_2016, DUTTA_2017, Ge_Pasquini_2017,Soumya_2018,ESSnuSB_1,Kaur_2021}. Constraints on non-unitary parameters are in refs. \cite{Giunti_2021,PARKE,Agarwalla:2021,Blennow_Pila:2016,Rahaman_2022,Ellis_Kelly:2020,Argawalla_2023,Moreno_2024,Soumya_2018,Escrihuela_2017,Miranda_2021, kozynets_2024}.

Motivated by these possible anomalies in previous experimental data and recognizing a promising opportunity to probe new physics with future experiments, we consider the possibility that the PMNS mixing matrix be a non-unitary matrix and
investigate its potential impact on the sensitivity to the CP-violating phase. Furthermore, we will assume different scenarios for non-unitarity, in low- and high-mass scales, aiming to categorize them based on their effects on the sensitivity of the CP phase. Our study differs from most of the previous work in the following ways:
\begin{itemize}
    \item We explicitly simulate near detectors, since, depending on the mass scale of new particles, the near detector plays a fundamental role in the sensitivity of the non-unitarity parameters. Furthermore, the impact of non-unitarity already in the near detector may influence the sensitivity of the standard oscillation parameters;
    \item aiming to get closer to reality, we also include the cross section uncertainty in our analysis, since the impact of this systematic error is extremely detrimental to CP sensitivity \cite{PilarHubber,Huber_2007-1}.
\end{itemize}

 The article is organized as follows: In Section \ref{sec:theory} we briefly discuss the non-unitarity approach and the parametrization adopted. Section \ref{sec:RelevantExp} describes the technical details of the experiments adopted and the procedures to carrying out the minimization. Section \ref{sec:SensNonUniPara} shows the sensitivity result for non-unitary parameters. In Section \ref{sec:CPSens} we give the results for the CP violation sensitivity over impact of non-unitary mixing taking into account the proposed situations for non-unitarity. We describe our conclusions in section \ref{sec:conclusion}. 

\section{Non-Unitary in Neutrino Oscillation}\label{sec:theory}
In the 3$\nu$ approach, the neutrino flavor states $\nu_{\alpha}$ ($\alpha = e,\mu,\tau$) is written as a linear combination of mass states $\nu_{i}$ ($i = 1, 2, 3$) weighted by a $3\times 3$ unitary mixing matrix $U$, in other words,
$\nu_{\alpha} = \sum_{i} U_{\alpha i} \nu_{i}$. The equation that govern the neutrino propagation in matter is given, in the mass basis, by
\begin{equation}
   i \frac{d \nu_{i}}{dx}  =  H \nu_{i} = (\mathcal{M} + U^{\dagger}\mathcal{A}U)\nu_{i} ,  
 \label{eq:propagation}  
\end{equation}
with $H$ being the Hamiltonian in the mass basis, $\mathcal{M}$ the matrix of mass eigenvalues and $\mathcal{A}$ the matter potential matrix.  The solution of Equation (\ref{eq:propagation}) enables us to comprehend the behavior of neutrinos as they propagate a distance $x$ through matter. This solution can be expressed as.
\begin{equation}
 \nu_{i}(x) = S_{i}(x)\nu_{i} =  (X e^{-i\Tilde{h}x}X^{\dagger})\nu_{i},
\end{equation}
where $S_{i}(x)$ is a matrix that describes possible changes in the mass basis. The transformation between the vacuum mass basis, $\nu_{i}$, and the matter mass basis, $\Tilde{\nu}_{i}$, is done by the unitary matrix $X$, such that $\nu_{i} = X\Tilde{\nu}_{i}$. The eigenvalues in the matter mass basis are $\Tilde{h} =\text{diag}(\Tilde{h}_{1}, \Tilde{h}_{2}, \Tilde{h}_{3}$). Unitary property of $U$, enable us to write $S$-matrix in the flavor basis, 
\begin{equation}
  S_{\beta \alpha}(x) = U S_{i}(x)U^{\dagger} = (UX)e^{-i\Tilde{h}x}(UX)^{\dagger}. \label{eq:smatrix}  
\end{equation}
where the transition between flavor is $\nu_{\beta}(x) = S_{\beta \alpha} \nu_{\alpha}$. Therefore, the probability to find a neutrino, initially produced as a flavor $\alpha$, and after traveled a distance $x$, as a flavor $\beta$ is 
\begin{equation}
P(\nu_{\alpha} \rightarrow \nu_{\beta}; x) = |S_{\beta \alpha}(x)|^{2}. \label{eq:prob}
\end{equation}

One can extend this formalism to a $(3 + n)$ unitary space with a unitary mixing matrix $\mathbf{U}$, where we have three active neutrino plus $n$ heavy neutrinos, called sterile neutrinos \cite{HIROSHI,Fong_2019}. In this case,
\begin{equation}
 \mathbf{U}^{(3 +n)\times(3 +n)} = 
 \begin{pmatrix}
  N & W  \\
  Z & V \\
 \end{pmatrix},
 \label{TotalMatrix}
\end{equation}
where $N_{3\times 3}$ is a non-unitary mixing matrix in the active space, space of well-know flavors $\nu_{\alpha} = e, \mu, \tau$. $W_{3\times n}$ and $Z_{n\times 3}$ are matrices in the active-sterile and sterile-active space, respectively, and $V_{n\times n}$ a matrix in the sterile-sterile space. We can use equations (\ref{eq:smatrix}) and (\ref{eq:prob}), just replacing the standard mixing matrix, $U$, with the new mixing matrix, $\mathbf{U}^{(3 +n)\times(3 +n)}$, to obtain the transition probability.   

The masses of these $n$ new particles allow us to separate the cause of unitarity violation in physics beyond the SM at high- and low- energy scales. In high-energy scales, one of the most salient features is that the mass of these new leptons is much heavier than the active neutrinos, therefore, they are not produced in the same process as neutrinos are produced, and the transition to them is kinematically forbidden. For general bounds on non-unitarity, in the context of high-scale unitarity violation, see refs. \cite{Antusch_2006, ESCRIHUELA,deGouvea:2015, Fernandez-Martinez:2016}. On the other hand, if new physics below the electroweak scale causes unitarity violation, in addition to mixing with neutrinos, these new particles participate in the phenomenon of neutrino oscillation. Here, it is important to note that the matrix (\ref{TotalMatrix}) is independent of the mass scale, in other worlds, if the extra states are kinematically accessible or not, the matrix is the same. 

In the low-energy scale, we need to distinguish between two cases. The first one is the case where the oscillations induced by extra heavier states are averaged out already at the near detector (ND). Therefore, ND can also be used to place limits on non-unitary parameters. We will refer to this case as \textbf{ND developed}. Considering the experimental setup for Hyper-Kamiokande, with a peak energy $\sim$ 0.6 GeV and a near detector baseline $\sim$ 1 km, the new mass is $\Delta m^{2} \gtrsim$ 10 eV$^{2}$. For DUNE, the peak energy is $\sim$ 2.5 GeV and the near detector distance is $\sim$ 0.5 km; therefore, this is the case where $\Delta m^{2} \gtrsim$ 100 eV$^{2}$.

In the second case, the oscillations induced by extra heavier states are averaged out only at the far detector, and the near detector measures the neutrino fluxes and cross-section predicted by the SM, without any indication of unitarity violation. We will refer to this case as \textbf{ND undeveloped}. Considering the experimental setup for Hyper-Kamiokande and DUNE, this is the case when 0.1 eV$^{2}$ $\lesssim \Delta m^{2} \lesssim$ 1 eV$^{2}$. 

For the two cases mentioned above on the low-energy scale, if the heavier neutrino masses are in the region $0.1$ eV$^{2}$ $\lesssim m^{2}_{J} \lesssim 1$ MeV$^{2}$, the model of unitarity violation is insensitive to details of the sterile neutrino sector \cite{HIROSHI}. Therefore, the matrix in (\ref{TotalMatrix}) becomes $\mathbf{U}^{(3 +n)\times(3 +n)} \rightarrow N_{3 \times 3}$. However, if $\Delta m^{2} \sim$ atmospheric or solar mass, there is no reason to make the model insensitive to the mass of the new particles, and severe model dependence must be expected when fitting data. This case will not be treated in this paper. 

In summary, we will simulate both cases without distinguishing between high- and low-mass scales of unitarity violation. In other words, we will not consider whether neutrinos are, or are not, kinematically accessible. We will simply categorize the cases as ``unitarity violation appears at both the near and far detector'' (ND developed) and ``unitarity violation appears only at the far detector'' (ND undeveloped), with ND developed potentially encompassing both low- and high- mass scales. Therefore, let us examine how each case affects the sensitivity of the non-unitary parameters and the CP sensitivity. Simulation details are given in section \ref{sec:simu}.   

\subsection{Parameterization}
In order to study the effects of the non-unitarity mixing characterized by the $3\times 3$ matrix $N$ in neutrino oscillation experiments, we follow the parameterization
considered in refs.  \cite{ESCRIHUELA, Soumya:2018nkw,Blennow_Pila:2016, Ellis_Kelly:2020}, so we have,
\begin{equation}
 N = 
\begin{bmatrix}
\xi_{11} & 0 & 0 \\
\xi_{21} & \xi_{22} & 0 \\
\xi_{31} & \xi_{32} & \xi_{33} \\
\end{bmatrix}
\times
U_{PMNS}, \label{eq:nonuni}  
\end{equation}
where $U_{PMNS}$ is the standard $3\times3$ unitary mixing matrix and $N$ is a non-unitary matrix if the triangular matrix $\xi$ is not 1; here, unitarity is recovered in the limit when $\xi_{ij} \rightarrow \delta_{ij}$. The diagonal elements $\xi_{ii}$ are real numbers, the off-diagonal elements $\xi_{ij}$ are complex numbers, and one can express them in terms of absolute values $|\xi_{ij}|$, and their phases $\phi_{ij}$, which can introduce a new CP phase. Here, the non-unitarity hypothesis is induced by $n$ new states, so-called sterile neutrinos, which are mixed with the active neutrinos \cite{HIROSHI, PARKE}. In this subclass of unitarity violation, row and columns normalization equations never exceeds 1, this can be seen directly from unitary condition \cite{Fong_2019},
\begin{equation}
\mathbf{U} \mathbf{U}^{\dagger} = I_{(3\times n)(n\times 3)}
\longrightarrow NN^{\dagger} + WW^{\dagger} = 1_{3\times 3}.
\end{equation}

Furthermore, one can invoke Cauchy-Schwarz inequalities in the mixing matrix elements \cite{PARKE}, to further bound the parameter space,
\begin{equation}
\left| \sum_{k}^{3} N_{\alpha k}N_{\beta k}^{*}\right|^{2} \leq \left(1 - \sum_{k}^{3} |N_{\alpha k}|^{2}\right)\left(1 - \sum_{k}^{3}|N_{\beta k}|^{2}\right), \hspace{4mm} \text{for} \hspace{4mm} \alpha = e, \mu, \tau, \hspace{4mm} \alpha \neq \beta,
\end{equation} 
in terms of the parametrization (\ref{eq:nonuni}) of $N$, Cauchy-Schwarz inequalities can relate the off-diagonal parameters to the diagonal parameters in the following way \cite{Giunti_2021,Ellis_Kelly:2020,PARKE},
\begin{equation}
|\xi_{ij}| \leq \sqrt{(1 - \xi_{ii}^{2})(1 - \xi_{jj}^{2})}. \label{eq:CS}  
\end{equation}

The equation (\ref{eq:CS}) tell us that we can constraint the off-diagonal non-unitary parameters only with information over the diagonal elements of the matrix. Furthermore, strong bounds in one of them, $\xi_{ii}$ or $\xi_{jj}$, is enough to constraint $|\xi_{ij}|$.

Using the non-unitary approach, we can write the new neutrino oscillation probability. For approximated analytic expressions of oscillation probabilities in matter with non-unitary effects,
see refs. \cite{Fong_2019, Fong:2022oim, Li:2015oal, Parke:2019jyu}. In vacuum we have,     
\begin{multline}
P(\nu_{\alpha} \rightarrow \nu_{\beta}) = \left| \sum_{k} N_{\alpha k} N^{*}_{\beta k}\right|^{2} -  4 \sum_{k>j} \text{Re}[ N_{\beta k}N^{*}_{\alpha k} N^{*}_{\beta j} N_{\alpha j}] \sin^{2} \left(\frac{\Delta_{kj}}{2}\right) + \\ 2 \sum_{k>j} \text{Im} [  N_{\beta k} N^{*}_{\alpha k} N^{*}_{\beta j} N_{\alpha j}] \sin \left(\Delta_{kj}\right).\label{osc_prob}
\end{multline}
where $\Delta_{kj} = \Delta m_{kj}^{2}L/2E$ with $\Delta m_{kj}^{2} = m_{k}^{2} - m_{j}^{2}$, $L$ is the distance between source and detector and $E$ is the neutrino energy. In terms of non-unitary parameters, the approximation to the electron neutrino appearance channel is given by \cite{ESCRIHUELA}
\begin{equation}
P_{\mu e} = (\xi_{11}\xi_{22})^{2}P_{\mu e}^{3\times 3} + \xi^{2}_{11} \xi_{22} |\xi_{21}| P_{\mu e }^{I} + \xi_{11}^{2}|\xi_{21}|^{2},
\label{eq:p_mue}
\end{equation}
where $P_{\mu e}^{3\times 3}$ is the standard oscillation probability depending only on the standard oscillation parameters. $P_{\mu e }^{I}$ is the term that contains the new CP phase $\phi_{21}$; here it is worth mentioning that this phase is related to the parameter $|\xi_{21}|$ and can lead to a degeneracy in the conversion probability. Explicitly, the terms $P_{\mu e}^{3\times 3}$ and $P_{\mu e }^{I}$ are given by \cite{Nunokawa_2008,ESCRIHUELA},
\begin{eqnarray}
P_{\mu e}^{3\times 3} &=& 4\left[ \cos^{2}\theta_{13} \cos^{2}\theta_{12} \cos^{2}\theta_{23}\sin^{2}\theta_{12}\sin^{2}\left(\frac{\Delta_{21}}{2}\right)
 + \cos^{2}\theta_{13}\sin^{2}\theta_{13}\sin^{2}\theta_{23} \sin^{2}\left(\frac{\Delta_{31}}{2}\right) \right]  \nonumber \\
 &+& \sin(2\theta_{21})\sin(2\theta_{23})\sin(2\theta_{13})\cos\theta_{13}\sin\left(\frac{\Delta_{21}}{2}\right)\sin\left(\frac{\Delta_{31}}{2}\right)\cos \left(\frac{\Delta_{31}}{2} + \delta_{CP}\right) 
 \label{eq:PueSTD}
\end{eqnarray}
and
\begin{eqnarray}
P_{\mu e }^{I} &=& -2\sin(2\theta_{13})\sin\theta_{23}\sin \left(\frac{\Delta_{31}}{2}\right) \sin \left(\frac{\Delta_{31}}{2} + \phi_{NP} + \delta_{CP}\right) \nonumber \\
&-& \cos\theta_{13}\cos\theta_{23}\sin(2\theta_{12})\cos\Delta_{21} \sin (\phi_{NP}),
\label{eq:PueI}
\end{eqnarray}
with $-\delta_{CP}\footnote{The $\eta_{ij}$ phases are Majorana phases. The ``invariant'' combination $\delta \equiv \eta_{12} - \eta_{13} + \eta_{23}$, corresponding to the ``Dirac phase'', exist only beyond three neutrinos. In the regime with three neutrinos, a single phase (say $\eta_{13}$) may be taken non-zero. This is the phase that corresponds to the Dirac CP phase in the quark sector, and affects neutrino oscillations involving three neutrinos \cite{Nunokawa_2008}.} = \eta_{12} - \eta_{13} + \eta_{23}$ and  $\phi_{NP} = \eta_{12} - \mathrm{Arg}(\xi_{21})$ being the new phase related to unitarity violation. In our analysis, we set $\eta_{12} = \eta_{23}$ = 0. For survival probability, we have the approximation \cite{ESCRIHUELA},
\begin{equation}\label{eq:P_mumu}
 P_{\mu \mu} = \xi_{22}^{4}P_{\mu \mu}^{3\times 3} + \xi_{22}^{3}|\xi_{21}|P_{\mu \mu}^{I_{1}} + 2\xi_{22}^{2}|\xi_{21}|^{2}P_{\mu \mu}^{I_{2}},      
\end{equation}
with $P_{\mu \mu}^{3\times 3}$ being the standard disappearance probability, which an approximate expression is given by,
\begin{align}
P_{\mu \mu}^{3\times 3} &\approx 1 
- \big[\sin^{2}(2\theta_{23})\cos^{2}\theta_{13} 
+ \sin^{4}\theta_{23}\sin^{2}(2\theta_{13}) \big]
\sin^{2}\left(\frac{\Delta_{31}}{2}\right) \nonumber \\
&\quad + \Bigg\{
\sin^{2}(2\theta_{23})\sin^{2}\theta_{13}
\Bigg[1 - \sin^{2}(2\theta_{12})
\left(\frac{1}{2} + \cos^{2}\delta_{\text{CP}}\right)\Bigg] \nonumber \\
&\quad + \sin(2\theta_{12})\cos^{2}\theta_{23}
\Bigg[\sin(2\theta_{12})\cos^{2}\theta_{23} \nonumber \\
&\quad + 2\cos(2\theta_{12})\sin(2\theta_{23})\sin(\theta_{13})\cos\delta_{\text{CP}}\Bigg]
\Bigg\}
\sin^{2}\left(\frac{\Delta_{21}}{2}\right).
\end{align}

The terms $P_{\mu \mu}^{I_{1}}$ and $P_{\mu \mu}^{I_{2}}$ are given by \cite{ESCRIHUELA},
\begin{eqnarray}
P_{\mu \mu}^{I_{1}}  & \approx &  -8\left[ \sin\theta_{13}\sin\theta_{23}\cos(2\theta_{23})\cos(\delta_{CP} +\phi_{NP})\right] \sin^{2}\left(\frac{\Delta_{31}}{2}\right) \nonumber \\
&+& 2\left[\cos\theta_{23}\sin(2\theta_{12})\sin^{2}\theta_{23}\cos(\phi_{NP})\right]\sin\Delta_{31} \sin\Delta_{21},
\end{eqnarray}
and
\begin{equation}
P_{\mu \mu}^{I_{2}} \approx 1 - 2\sin^{2}\theta_{23}\sin^{2}\left(\frac{\Delta_{31}}{2}\right).
\end{equation}

The exact equations for muon neutrino appearance and disappearance in vacuum, assuming the three-neutrino flavor scenario, are given in Appendix \ref{AppendixB}. We can see that in vacuum in the presence of non-unitarity, appearance and disappearance channels are mainly driven by $\xi_{11}$, $\xi_{22}$ and $|\xi_{21}|$, therefore, an experiment, in which vacuum is a good approximation, can put strong limits in these non-unitary parameters. The remaining non-unitary parameters, $\xi_{33}$, $|\xi_{31}|$ and $|\xi_{32}|$ only appear in the presence of matter effects \cite{ESCRIHUELA, Agarwalla:2021}. In our analysis, the results are obtained by numerical computations in the presence of matter effects without using any approximated analytic expressions such as those performed in \cite{HIROSHI}. Furthermore, within the probability equations presented in ref. \cite{HIROSHI}, there appears what has been designated as ``the leaking term'', a term which denotes non-unitary violation models at low scales; in our analysis, we are ignoring this term as it is expected to be very small.   

\begin{figure}[ht]
\centering
\includegraphics[scale=0.80]{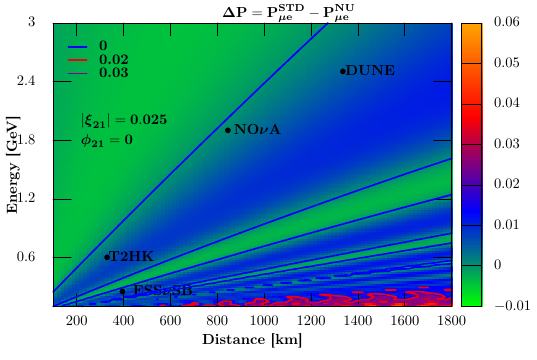}
\includegraphics[scale=0.80]{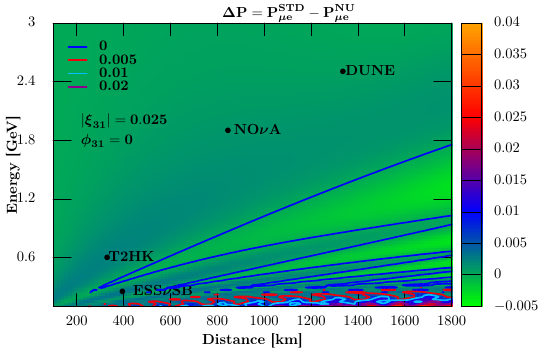}
\caption{Iso-contour of difference $\Delta P$ in the $L-E$ plane. The left (right) panel represents the deviation from unitary due to $\xi_{21}$ ($\xi_{31}$). See text for details.}  \label{fig:map_probability}
\end{figure}

In order to show the impact of non-unitarity parameters, Figure \ref{fig:map_probability} shows the deviation of non-unitary from unitary mixing as a function of the possible baseline and neutrino energy. The difference is defined as $\Delta P = P_{\mu e}^{\text{STD}} - P_{\mu e}^{\text{NU}}$, where $P_{\mu e}^{\text{STD}}$ and $P_{\mu e}^{\text{NU}}$ are the probability in the unitary and non-unitary schemes, respectively. The standard and non-unitary parameters used are described in Section \ref{sec:simu}. The left (right) panel represents the deviation from unitary due to $\xi_{21}$ ($\xi_{31}$) parameter. Some long baseline neutrino experiments (LBNE) are shown for comparison. In Figure \ref{fig:map_probability}, we can see that those LBNE \cite{T2HK, T2HKK,DUNE-DESING, Baussan_2014,NOVA, NOVA_1} have a relation $L/E$ where non-unitarity starts to become relevant, which means that they could be sensitive to this scenario. 

\subsection{CP Asymmetry and Bi-Probability}
The magnitude of charge-parity (CP)-violation in neutrino oscillation can be described by the difference between neutrino and antineutrino channels. In the unitary case, in vacuum, we can write \cite{Nunokawa_2008}, 
\begin{equation}
P(\nu_{\alpha} \rightarrow \nu_{\beta}) - P(\bar{\nu}_{\alpha} \rightarrow \bar{\nu}_{\beta}) = - 16J_{\alpha \beta} \sin \left(\frac{\Delta_{21}}{2}\right)\sin \left(\frac{\Delta_{32}}{2}\right) \sin \left(\frac{\Delta_{31}}{2}\right),
\label{eq:PuePaue}
\end{equation}
where $J_{\alpha \beta}$ is the Jarlskog invariant, invariant that  controls the size of CP violation in the lepton sector. Using the standard parametrization of PMNS matrix, the invariant is written as
\begin{equation}
    J_{\alpha \beta} \equiv \mathrm{Im}[U_{\alpha 1} U_{\alpha 1}^{*}U_{\beta 1}^{*}U_{\beta 2}] = \pm J, \hspace{3mm} J \equiv s_{12} c_{12} s_{23}c_{23}s_{13}c^{2}_{13} \sin \delta_{CP},
\end{equation}

with positive and negative signs for cyclic and anti-cyclic permutations of the flavor indices $e$, $\mu$, and $\tau$, respectively. From equation (\ref{eq:PuePaue}), it is clear that there is no CP violation if $\delta_{CP}$ = 0 or $\pi$ and if at least one of the mixing angles is zero. In matter, assuming first-order in matter effect, we can consider as a measure of CP violation, the following neutrino-antineutrino probability asymmetry, which is approximately given by, in the unitary scenario \cite{Minakata_2001}, 
\begin{eqnarray}
A_{CP} = \frac{P(\nu_{\alpha} \rightarrow \nu_{\beta}) - P(\bar{\nu}_{\alpha} \rightarrow \bar{\nu}_{\beta})}{P(\nu_{\alpha} \rightarrow \nu_{\beta}) + P(\bar{\nu}_{\alpha} \rightarrow \bar{\nu}_{\beta})},
\label{eq:AcpUnitary}
\end{eqnarray}
with
\begin{eqnarray}
P(\nu_{\alpha} \rightarrow \nu_{\beta}) & -& P(\bar{\nu}_{\alpha} \rightarrow \bar{\nu}_{\beta}) \simeq -16 J_{\alpha \beta} \sin \left(\frac{\Delta_{21}}{2}\right)\sin \left(\frac{\Delta_{32}}{2}\right)\sin \left(\frac{\Delta_{31}}{2}\right)  \nonumber \\
&+& 2 \cos (2\theta_{13})\sin^{2}(2\theta_{13})\sin^{2}\theta_{23}\left(\frac{2 E a}{\Delta m^{2}_{31}}\right)\sin^{2}\left(\frac{\Delta_{31}}{2} \right) \nonumber \\
&-& \frac{aL}{2}\sin^{2}(2\theta_{13})\cos(2\theta_{13})\sin^{2}\theta_{23}\sin \Delta_{31}, 
\label{eq:AcpUnitary1}
\end{eqnarray}
and
\begin{align}
P(\nu_{\alpha} \rightarrow \nu_{\beta}) + P(\bar{\nu}_{\alpha} \rightarrow \bar{\nu}_{\beta}) \simeq 2\sin^{2}(2\theta_{13})\sin^{2}\theta_{23}\sin^{2}\left(\frac{\Delta_{31}}{2}\right) \nonumber \\
\quad - \sin^{2}\theta_{12} \sin^{2}(2\theta_{13})\sin^{2}\theta_{23}\sin \Delta_{21}\sin \Delta_{31} + 4 J_{\alpha \beta} \operatorname{cotg} \delta_{CP}\sin \Delta_{21}\sin \Delta_{31},
\label{eq:AcpUnitary2}
\end{align}
with $a = \sqrt{2} G_{F}N_{e}$ where $G_{F}$ is the Fermi constant and $N_{e}$ denotes the electron number density in the earth. Equations (\ref{eq:AcpUnitary1})-(\ref{eq:AcpUnitary2}) highlights the effects of matter on neutrino oscillations, even if $\delta_{CP}$ = 0 or $\pi$, $A_{CP}$ is non-zero, which introduces a false CP violation effect. Therefore, for long-baseline neutrino experiments where the matter effects are not negligible, the genuine leptonic CP violation must be disentangled from the matter effects.    

Now, consider a scenario with unitarity violation. Using equation (\ref{eq:p_mue}), the neutrino-antineutrino asymmetry in vacuum can be expressed as follows,
\begin{equation}
A_{CP} = \frac{(\xi_{11}\xi_{22})^{2}[P_{\mu e}^{3\times 3} - \bar{P}_{\mu e}^{3\times 3}] + \xi_{11}^{2}\xi_{22}|\xi_{21}|[P_{\mu e}^{I} - \bar{P}_{\mu e}^{I}]}{(\xi_{11}\xi_{22})^{2}[P_{\mu e}^{3\times 3} + \bar{P}_{\mu e}^{3\times 3}] + \xi_{11}^{2}\xi_{22}|\xi_{21}|[P_{\mu e}^{I} + \bar{P}_{\mu e}^{I}] + 2\xi_{11}^{2}|\xi_{21}|^{2}}.
\label{eq:AcpUnitaryII}
\end{equation}
Replacing equations (\ref{eq:PueSTD}) and (\ref{eq:PueI}), numerator and denominator of equation (\ref{eq:AcpUnitaryII}), can be written, respectively,
\begin{align}
\approx  & (\xi_{11}\xi_{22})^{2}\Bigg[-2\sin(2\theta_{12})\sin(2\theta_{23})\sin(2\theta_{13})\cos\theta_{13}\sin\left(\frac{\Delta_{21}}{2}\right) \sin^{2}\left(\frac{\Delta_{31}}{2}\right) \sin \delta_{CP}\Bigg] \nonumber \\
& - \xi_{11}^{2}\xi_{22}|\xi_{21}|\Bigg[4\sin(2\theta_{13})\sin \theta_{23} \sin \left(\frac{\Delta_{31}}{2}\right) \cos\left(\frac{\Delta_{31}}{2} + \phi_{NP}\right) \sin \delta_{CP} \Bigg]
\label{eq:numeratoAcp}
\end{align}

and 
\begin{align}
\approx & (\xi_{11}\xi_{22})^{2}\Bigg[ 8\left(\cos^{2}\theta_{13}\cos^{2}\theta_{12} \cos^{2}\theta_{23}\sin^{2}\theta_{12}\sin^{2}\left(\frac{\Delta_{21}}{2}\right) + \cos^{2}\theta_{13} \sin^{2}\theta_{13}\sin^{2}\theta_{23}\sin^{2}\left(\frac{\Delta_{31}}{2}\right)\right)   \nonumber \\
&+  2\sin (2\theta_{12})\sin (2\theta_{13})\sin (2\theta_{23})\cos\theta_{13}\sin\left(\frac{\Delta_{21}}{2}\right)\sin\left(\frac{\Delta_{31}}{2}\right) \cos\left(\frac{\Delta_{31}}{2}\right)\cos \delta_{CP}  \Bigg] \nonumber \\
& - \xi_{11}^{2}\xi_{22}|\xi_{21}|\Bigg[4\sin(2\theta_{13})\sin \theta_{23} \sin \left(\frac{\Delta_{31}}{2}\right) \sin\left(\frac{\Delta_{31}}{2} + \phi_{NP}\right)\cos\delta_{CP}   \nonumber \\  
&-2\cos\theta_{13}\cos\theta_{23}\sin(2\theta_{12})\cos\Delta_{21}\sin\phi_{\mathrm{NP}} \Bigg] + 2\xi_{11}^{2}|\xi_{21}|^{2}.
\label{eq:denominatorAcp}
\end{align}

\begin{figure}[ht]
\centering
\includegraphics[scale=0.80]{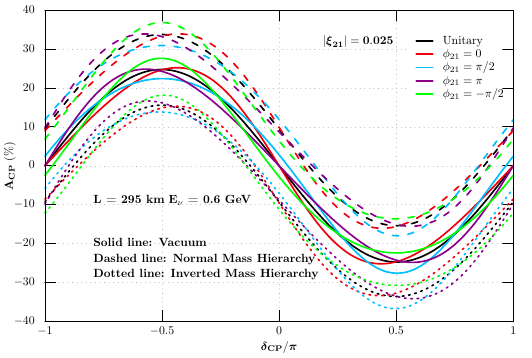}
\includegraphics[scale=0.80]{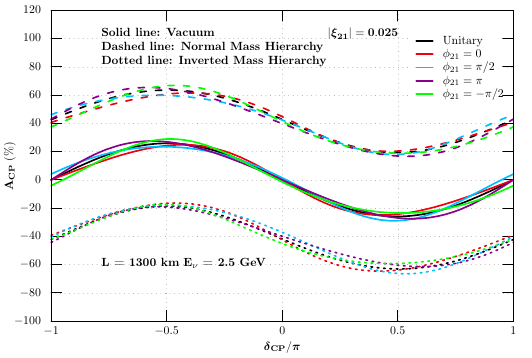}
\caption{The neutrino-antineutrino asymmetry as a function of $\delta_{CP}$ values for $L = 295$ km and $E_{\nu} = 0.6$ GeV (left) and $L = 1300$ km and $E_{\nu} = 2.5$ GeV (right). Unitary case is represented by black lines (solid, dashed and dotted). Non-unitary case is represented by colored lines (solid, dashed and dotted) with $\xi_{21} = 0.025$ and four values for $\phi_{21}$ (see legend).}
\label{fig:assimetry}
\end{figure}

From equations (\ref{eq:numeratoAcp}) and (\ref{eq:denominatorAcp}), it can be observed that a new phase, $\phi_{21}$, is introduced which has the potential to mimic the standard CP phase and enhance asymmetry. Figure \ref{fig:assimetry} shows how CP-violation, mass ordering and unitarity violation affect the difference between $\nu_{\mu} \rightarrow \nu_{e}$ detection probability relative to $\bar{\nu}_{\mu} \rightarrow \bar{\nu}_{e}$ detection probability for the set of parameters described in Section \ref{sec:simu}. 

We can see on the left side of Figure \ref{fig:assimetry} that the lack of knowledge about neutrino mass ordering affects the sensitivity to the CP phase, since we cannot distinguish these two cases (normal or inverted ordering) due to the small matter effect. On the right side, we have the opposite. For the DUNE distance, corresponding to a longer baseline, the asymmetry is more sensitive to the matter effect, which means that both the mass ordering and $\delta_{CP}$ can be determined unambiguously. It seems that the relatively small amount of non-unitarity does not induce a substantial enhancement in parameter degeneracy, not causing any significant reduction in sensitivity to CP violation. Furthermore, Figure \ref{fig:assimetry} shows a non-zero asymmetry even when $\delta_{CP} = 0, \pi$ and in vacuum, suggesting that the asymmetry equation contains a term dependent solely on $\sin(\phi_{NP})$. Our equations (\ref{eq:numeratoAcp}) and (\ref{eq:denominatorAcp}) do not explicitly exhibit this dependence, this is attributable to the approximate probability equations employed; we are following ref. \cite{ESCRIHUELA}. However, utilizing the approximate appearance equations from ref. \cite{Ge_Pasquini_2017} readily reveals this dependence, consistent with our result obtained in Figure \ref{fig:assimetry}.

\begin{figure}[ht]
\centering
\includegraphics[scale=0.43]{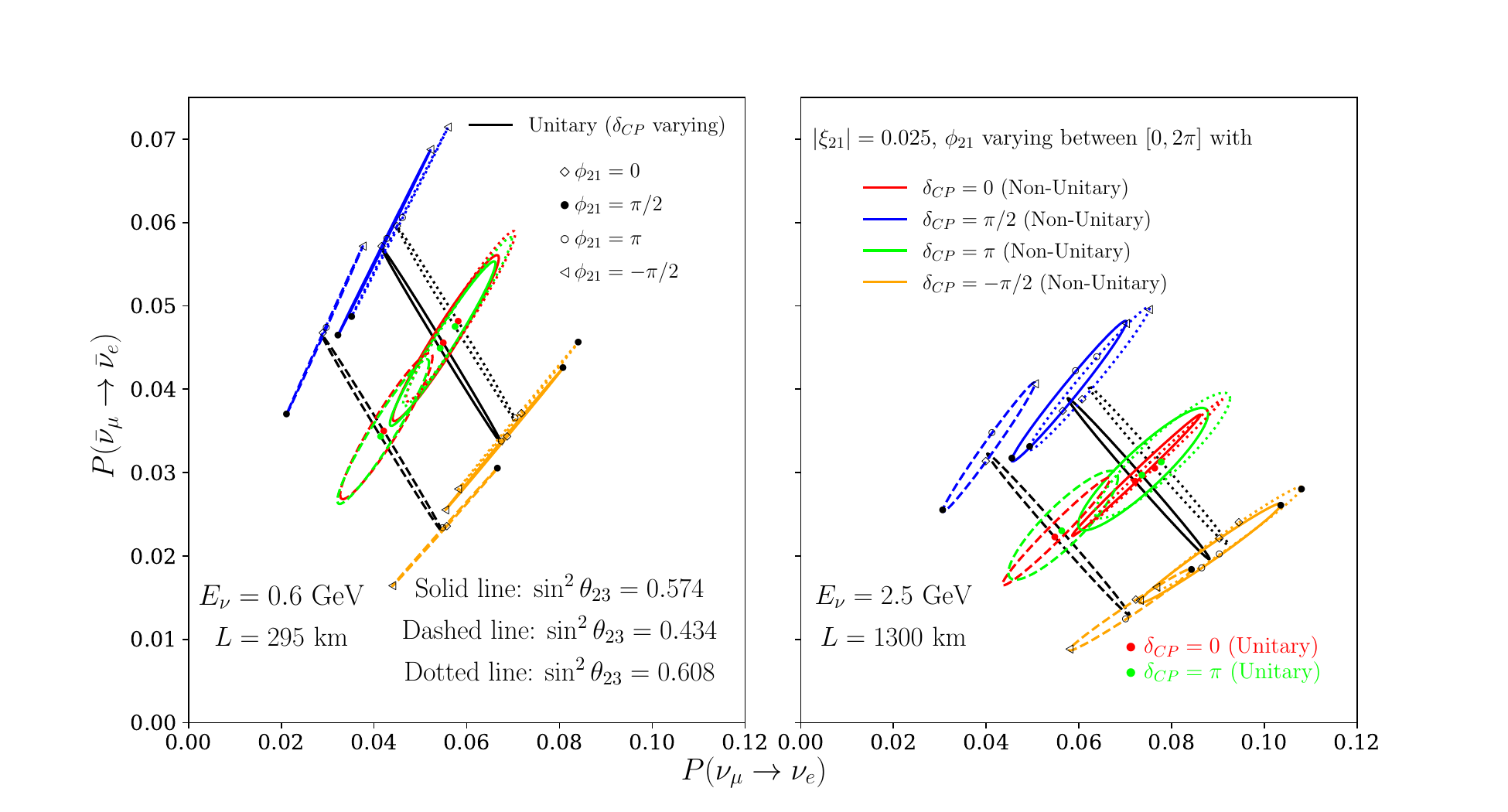}
\caption{CP ellipses in the bi-probability plane for unitary scenario with varying $\delta_{CP}$ (black line) and unitarity violation scenario with $\delta_{CP}$ fixed and varying $\phi_{21}$ (colored lines) for a setup with $E_{\nu} = 0.6$ GeV - $L = 295$ km (left) and $E_{\nu} = 2.5$ GeV - $L = 1300$ km (right). The solid lines correspond to $\sin^{2}\theta_{23} = 0.574$, dashed lines to $\sin^{2}\theta_{23} = 0.434$ and dotted lines to $\sin^{2}\theta_{23} = 0.608$.}
\label{fig:biprob}
\end{figure}

In addition to CP asymmetry, complementary information can be obtained from bi-probability plot. In Figure \ref{fig:biprob}, we show the CP ellipses for $L = 295$ km and $E_{\nu} = 0.6$ GeV (left) and $L = 1300$ km and $E_{\nu} = 2.5$ GeV (right). Normal mass ordering was assumed. The unitary case is represented by black lines where $\delta_{CP}$ was varied in the range $[-\pi, \pi]$. The non-unitary case is represented by colored lines, where $\delta_{CP}$ is fixed at a value (see the legends in the figure) and $\phi_{21}$ is varying in the range $[-\pi, \pi]$. For the solid line, we set $\sin^{2}\theta_{23} = 0.574$, dashed lines with $\sin^{2}\theta_{23} = 0.434$ and dotted lines with $\sin^{2}\theta_{23} = 0.608$ \cite{deSalasGlobalFit2020}.    

The following interesting features can be observed from Figure \ref{fig:biprob}:

\begin{itemize} 
    \item The combination of neutrino and antineutrino channels can break the degeneracy in the phases $\delta_{CP}$ and $\phi_{ij}$ for experiments situated in regions with a $L/E$ ratio close to 500 km/GeV, thus eliminating the impact of non-unitarity. This observation has been previously reported in reference \cite{Miranda_2016}. However, reference \cite{Miranda_2016} emphasizes that this is only valid in vacuum, and we aim to extend this observation to experiments where matter effects cannot be neglected. As shown in the right panel of Figure \ref{fig:biprob}, which also has a $L/E = 520$ km/GeV, the impact of non-unitarity does not significantly increase the degeneracy between $\delta_{CP}$ and $\phi_{21}$ phases.
    \item Considering the case where we aim to differentiate between $\delta_{CP} = \pm \pi/2$ from $\delta_{CP} = 0, \pi$, let us first assume that we are comparing the unitary case with the non-unitary case. For each point in the unitary case (black line), we compare it to the line corresponding to the colored case $\delta_{CP} = 0$ (red line) and $\delta_{CP} = \pi$ (green line). In our simulation this case will correspond to the scenario (a) $N^{\mathrm{true}}$ is computed assuming unitary. Now, we will compare non-unitary cases with non-unitary cases (only colored lines). In our simulation, these situations will correspond to the scenario\footnote{This situation will also correspond to the scenario (b) $N^{\mathrm{true}}$ is computed assuming non-unitary, only the value of $|\xi_{21}|$ is different but the discussions made here are also valid.} (c) $N^{\mathrm{true}}$ is computed assuming non-unitary. Note that the distance between points $\delta_{CP} = \pi/2,-\pi/2$ and $\delta_{CP} = 0, \pi$ is almost the same in both scenarios considered. In other words, we cannot distinguish between these cases; therefore, these different scenarios would yield the same impact on CP sensitivity.  
\end{itemize}

\section{Relevant Future Experiments}\label{sec:RelevantExp}
In this section, we briefly describe the experiments that will be considered in this paper. We focus, especially, on the Hyper-Kamiokande  \cite{T2HK,T2HKK} and DUNE \cite{DUNE} experiments. 

\subsection{Hyper-Kamiokande}
The Hyper-Kamiokande (HK) experiment consists of a water-Cherenkov detector, located in Kamioka, Japan, with a fiducial mass of 187 kton; approximately 8.4 times larger than the Super-Kamiokande \cite{HK-snowmass}. The neutrino beam is produced in the J-PARC proton synchrotron accelerator, located in Tokai, Japan, 295 km away from HK detector. We will refer to the Tokai to Kamioka setup in this paper as T2HK henceforth. The detector is designed for a wide variety of neutrino physics and astrophysics studies, nucleon decay searches, measuring neutrino oscillation parameters such as $\theta_{23}$, $\Delta m_{32}^{2}$, and in particular to determine the $\delta_{CP}$ phase are the main goals of the T2HK experiment \cite{T2HK}. The possibility of building an identical second Hyper-Kamiokande detector in Korea with a baseline around 1000-1300 km is under discussion \cite{T2HKK}. For simplicity, in this article we call this setup of having two identical detectors in Japan and Korea at the same time as T2HKK. T2HKK is expected to have a better ability to establish the existence (or not) of CP violation in the neutrino sector and would improve the precision of all oscillation parameters, allowing us to test the 3$\nu$ paradigm with better sensitivity. The near detector of the HK experiments considered in our analysis is the IWCD off-axis water Cherenkov detector, with a mass of 1 kt and located at a baseline of 1 km away from the source \cite{T2HK,hkNear,JWilson_2020}.

\subsection{DUNE}
The Deep Underground Neutrino Experiment (DUNE) is a long baseline neutrino experiment that utilizes a high-intensity neutrino beam produced at the Fermi National Accelerator Laboratory (Fermilab) in Batavia, Illinois. DUNE employs a two-detector system, the near detector (ND) is located at 574 m from the source at Fermilab and with 67 tons of Liquid Argon TPC (LArTPC) serves as a reference for understanding the properties of the neutrino beam before it travels long distances. The far detector (FD) is located deep underground at the Sanford Underground Research Laboratory in Lead, approximately 1300 kilometers away from the source and employs a massive liquid argon time-projection chamber (LArTPC) of 35-40 kton. \cite{DUNENear, Coloma_2021}. Among the broad physics and astrophysics program of the DUNE experiment, one of its primary goals is to measure the parameters governing neutrino oscillations with unprecedented precision, including the determination of the neutrino mass ordering and the search for possible CP violation in neutrino oscillations. These measurements could provide crucial insights into the fundamental nature of neutrinos and potentially shed light on the asymmetry between matter and antimatter in the universe. 

\subsection{Simulation Details}\label{sec:simu}
The simulation was performed with the GLoBES software \cite{Huber_2005,Huber_2007}, modifying the standard code to accommodate the non-unitary scenario. Our simulation of T2HK/T2HKK is based on description provided in refs. \cite{T2HK,T2HKK}. We also used the official GLoBES simulation files released by the DUNE Collaboration \cite{DUNEGlobes, DUNE-DESING}, which have the same experimental configurations as the DUNE TDR. In our analysis, we will simulate both near and far detectors; therefore, the final $\chi^{2}$ function implemented is given by

\begin{equation}
\chi_{\text{total}}^{2} =\underset{\vec{\eta}}{\mathrm{min}}  \left( \sum_{D}\sum_{k}^{\text{bins}} \frac{\left[N_{D, k}^{\text{true}}(\xi_{ij}) -N_{D, k}^{\text{fit}}(\xi_{ij}, \vec{\eta})\right]^{2}}{N_{D, k}^{\text{true}} (\xi_{ij})} + \sum_{n}\frac{\eta^{2}_{n}}{\sigma_{n}^{2}} \right)
\label{eq:chi_squareTotal}
\end{equation}  
where $N^{\text{true}}$ is the expected observed number of events, that can be computed without or with unitarity violation. $N^{\text{fit}}$ is the number of events simulated to test our non-unitarity scenario where non-unitary parameters are represented by $\xi_{ij}$. The sum over the detectors, near and far, is represented by the index $D$. Systematic uncertainties are represented by the vector $\vec{\eta}$, for both detectors with corresponding one sigma uncertainties $\sigma_{n}$. During analysis, both $\nu_{e}$ appearance and $\nu_{\mu}$ disappearance samples, in neutrino and anti-neutrino mode were used.

For the near and far detector, systematic uncertainties are normalization error, energy calibration error (tilt), and ratio of the electron neutrino to antineutrino cross-section. Throughout this article, we used, for both T2HK and T2HKK far detector, 2.5\% (2\%) normalization and 3\% energy calibration for $\nu_{e}$ ($\bar{\nu}_{e}$)-like events and 1\% normalization and 3\% energy calibration for $\nu_{\mu}$($\bar{\nu}_{\mu}$)-like events \cite{Xie:2023}. For the DUNE far detector, 2\% normalization for $\nu_{e}$ ($\bar{\nu}_{e}$)-like events and 5\% normalization for $\nu_{\mu}$($\bar{\nu}_{\mu}$)-like events were assumed \cite{DUNEGlobes}. For the near detector of T2HK/T2HKK, the normalization uncertainty is 10\% for the neutrino channel and 12\% for the antineutrino channel. The uncertainty values in the energy calibration (tilt) are the same as those of FD. For the near detector of DUNE, the normalization uncertainty is 6\% for $e$-like events and 15\% for $\mu$-like events. In the far detector for all the experiments considered, the uncertainty in the ratio of electron neutrino to antineutrino cross-section is set to $\sigma_{\nu /\bar{\nu}} = 4.9$\% at present the best constraint, and $\sigma_{\nu /\bar{\nu}} = 2.7$\% as future constraint \cite{Naseby:2021}. In the near detector, we implement $\sigma_{\nu /\bar{\nu}}$ = 8.5\% for all experiments. We have assumed that the systematic uncertainties in the near detector are larger than those in the far detector. Consequently, in our analysis, the uncertainties associated with the ND were scaled by a factor of two or three, depending on the specific uncertainty, relative to the uncertainties of the FD. Furthermore, for the sake of simplicity, identical signal and background rejection efficiencies, bin sizes, and energy resolutions were assumed for both near and far detectors. However, this assumption may not hold universally, and discrepancies could arise due to variations in detector technology or design. The standard oscillation parameters used during this analysis are shown in Table \ref{tab:std_osc} \cite{PDGNavas2024,deSalasGlobalFit2020}. Detailed information on the total event rates considered in our analysis is available in the Appendix \ref{AppendixA}. The solar and reactor parameters, as well as the atmospheric mass, $\Delta m_{31}^{2}$, are fixed in their best-fit values. We marginalize over $\theta_{23}$ ($3\sigma$ range) and $\delta_{CP}$ in [$-\pi, \pi$] range, with no penalty term.
Normal ordering was assumed during the entire paper, unless otherwise stated. The technical details of the experiments are shown in Table \ref{technical_info}.

\renewcommand\arraystretch{1.2}
\renewcommand\tabcolsep{0.2cm}
\begin{table}[ht]
\caption{Standard oscillation parameters used throughout the paper \cite{PDGNavas2024,deSalasGlobalFit2020}.}
\centering
\begin{tabular}{|c|c|c|c|c|c|}
\hline
$\sin^{2} \theta_{12}$ & $\sin^{2} \theta_{13}$ & $\sin^{2} \theta_{23}$ & $\delta_{CP}$ & $\Delta m_{21}^{2}$ & $\Delta m_{32}^{2}$ \\
\hline
0.307 & 0.0219 & 0.574 & -$\pi$/2 & 7.53$\times$10$^{-5}$ & 2.455$\times$10$^{-3}$ \\
\hline
fixed & fixed & marginalized & marginalized & fixed & fixed \\
\hline
\end{tabular}
\label{tab:std_osc}
\end{table}

\renewcommand\arraystretch{1.2}
\renewcommand\tabcolsep{0.2cm}
\begin{table}[ht]
\caption{Technical information of T2HK/T2HKK \cite{T2HK,T2HKK} and DUNE \cite{DUNE} used in our simulation.}
\centering
\begin{tabular}{|c|c|c|c|}
\hline
 & T2HK & T2HKK & DUNE \\
\hline
L(km)/E(GeV) & 295/0.6 & 295/0.6 + 1100/0.8 & 1300/2.5 \\
\hline
Detector Mass (kton) & 187 & 187(295 km) + 187(1100 km) & 40 \\
\hline
Beam Power (MW) & 1.2 & 1.2 & 1.3 \\
\hline
POT/year & 1.1$\times$ 10$^{21}$ &1.1$\times$ 10$^{21}$ & 2.7$\times$ 10$^{21}$  \\
\hline
Exposure Time ($\nu:\bar{\nu}$) & 2.5:7.5 & 2.5:7.5 & 3.5:3.5 \\
\hline
\end{tabular}
\label{technical_info}
\end{table}

Following the discussions in Section \ref{sec:theory}, depending on the heavier neutrino mass, unitarity violation may or may not manifest over short distances. We have defined two different cases: ND developed is when non-unitarity appears at short distance, and consequently near detectors may be useful to constraint non-unitarity parameters. ND undeveloped, when near detector measures the SM flux and cross section, and non-unitarity play a fundamental role only at far detector. It is important to verify how each case affects the sensitivity of the $\delta_{CP}$ phase, therefore, we will consider the following three situations for ND developed and ND undeveloped, and with the non-unitary phase $\phi_{21}$ assuming four values, 0, $\pi/2$, $\pi$ and $-\pi/2$,   
\begin{itemize}
\item[(a)]  $N^{\text{true}}$ is computed assuming unitarity, therefore, $|\xi_{ij}| = 0$, $\xi_{ii} = 1$.  
\item[(b)] $N^{\text{true}}$ is computed assuming non-unitarity with $|\xi_{21}|$ = 0.01, $\xi_{11}$ = 0.99, $\xi_{22}$ = 0.99 (intermediate case) \cite{Giunti_2021, Blennow_Pila:2016}. The other parameters are set $\xi_{31} = \xi_{32} = 0$ and $\xi_{33} = 1$.
\item[(c)] $N^{\text{true}}$ is computed assuming non-unitarity with $|\xi_{21}|$ = 0.025, $\xi_{11}$ = 0.975, $\xi_{22}$ = 0.99 (upper limit case) \cite{Giunti_2021, Blennow_Pila:2016}. The other parameters are set $\xi_{31} = \xi_{32} = 0$ and $\xi_{33} = 1$.
\end{itemize}

Our goal with the above situations is to verify how each case reduces the sensitivity and whether we can differentiate them. In addition to the already marginalized parameters, we also marginalize over non-unitary parameters, $\xi_{ii}$, $|\xi_{ij}|$ and $\phi_{ij}$, respecting Cauchy-Schwarz inequality, equation (\ref{eq:CS}). 

\section{Non-Unitary Parameters Sensitivity}\label{sec:SensNonUniPara}
In this section, we present the bounds on non-unitary parameters using long-baseline neutrino experiments in the presence of non-unitary neutrino mixing. First, we estimate the sensitivity to the non-unitary parameters for the case where $N^{\mathrm{true}}$ is computed assuming unitarity, for both scenarios, the ND developed and ND undeveloped. Later, by fixing $|\xi_{21}|^{\mathrm{true}} = 0.025$ we study the capability of T2HK, T2HKK and DUNE to constraint the dominant non-unitary parameters in the $\nu_{e}$-appearance channel, which are $|\xi_{21}|$ and $\phi_{21}$.

\subsection{Case where $N^{\mathrm{true}}$ is computed assuming unitarity}

\begin{figure}[ht]
\centering
\includegraphics[scale=0.82]{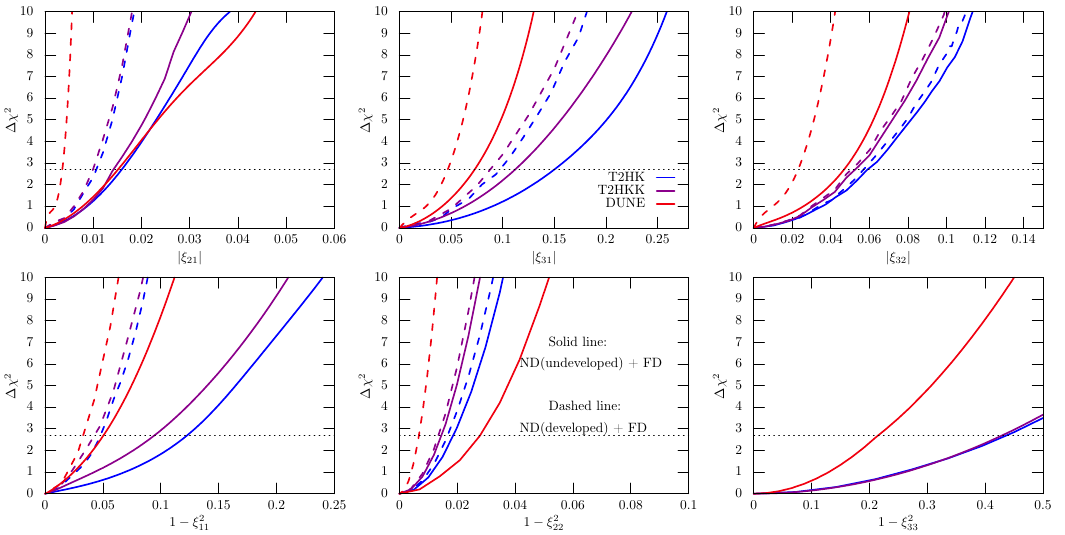}
\caption{Expected sensitivity on non-unitary parameters to be achieved by the T2HK, T2HKK and DUNE experiments. The upper panels show the sensitivity to the off-diagonal parameters and lower panels to the diagonal parameters. The case where non-unitarity only appears in the far detector (ND undeveloped) is represented by solid lines. The case where unitarity violation already appears in the near detector (ND developed) is represented by dashed lines. The dashed horizontal line represents the $\Delta \chi^2$ value corresponding to 90\% CL.}  \label{fig:projection}
\end{figure}

In Figure \ref{fig:projection}, we plot the expected sensitivity to different non-unitary parameters to be achieved by T2HK, T2HKK and DUNE experiments. The non-unitary represented in this figure are the cases corresponding to ND undeveloped (solid line) and ND developed (dashed line). The upper (lower) panels show the bounds that can be obtained on the off-diagonal (diagonal) parameters.

We can see in Figure \ref{fig:projection}, considering ND developed and ND undeveloped, all experiments can significantly constraint almost all non-unitary parameters, but $\xi_{33}$. The reason is that the $\xi_{33}$ parameter is related to the $\tau$-channel, which requires $\tau$ neutrino detection, and currently has poor statistics. Although $\xi_{33}$ appears in the probability equation linked with matter effects, information from $\tau$-channel is essential to constraint this parameter, as shown in ref. \cite{Giunti_2021}, in which was used neutral-current (NC) data from MINOS/MINOS+ and achieved a better limit compared to our results. Also, note that the limit on the parameter $\xi_{33}$ is unaffected by the kind of ND scenario used, as the probability for short distances is independent of $\xi_{33}$. 

The limits on the $\xi_{22}$ parameter come from the $\nu_{\mu} \rightarrow \nu_{\mu}$ channel, both in the vacuum and in the matter framework, where in the probability, $\xi_{22}$, is the dominant term; therefore, T2HK/T2HKK and DUNE can put strong limits on this parameter due to high statistics (see the middle-lower panel in Figure \ref{fig:projection}). The parameter $\xi_{11}$ is constrained mainly by the $\nu_{\mu} \rightarrow \nu_{e}$ channel, however, in this channel, due to the uncertainty of $\delta_{CP}$, we can observe a loss of sensitivity compared to $\xi_{22}$. Also, it is important to note that there is a clear difference between the limits obtained under the ND developed and the ND undeveloped scenario. In the ND developed scenario, there is a significant improvement in the sensitivity of both parameters, $\xi_{11}$ and $\xi_{22}$, for all considered experiments compared to the ND undeveloped case. For $\xi_{11}$, the improvement is largely due to the beam contamination channel, which is directly proportional to $\xi_{11}$ with $P_{ee} = \xi_{11}^{4}P_{ee}^{\mathrm{std}}$, and has a considerable impact on sensitivity.  The disappearance probability when $L \rightarrow 0$ is manly driven by $\xi_{22}$, thus leading to a significant improvement in sensitivity. 

Among the off-diagonal parameters, the best constrained one is $|\xi_{21}|$, in both the ND developed and the ND undeveloped scenarios. As $|\xi_{21}|$ is the main parameter in the $\nu_{\mu} \rightarrow \nu_{e}$ channel, due to the large statistics and good sensitivity, T2HK/T2HKK and DUNE can put a strong limit. For the remaining non-unitary parameters, $|\xi_{31}|$ and $|\xi_{32}|$, the sensitivity comes exclusively from the Cauchy-Schwarz inequality, equation (\ref{eq:CS}), mainly T2HK, which is weakly dependent on $|\xi_{31}|$ and $|\xi_{32}|$. Although T2HKK has dependence on $|\xi_{31}|$ and $|\xi_{32}|$ because of the matter effect, the limits for these parameters come from the appearance channel, in which T2HKK has a small number of events. 

The corresponding 90\% C.L. limits are summarized in Table \ref{tab:bounds}. Note that our results on non-diagonal of the non-unitary parameters, $|\xi_{31}|$ and $|\xi_{32}|$, are weaker than those in ref. \cite{Giunti_2021}. As previously mentioned, in \cite{Giunti_2021} the authors used neutral current data from the MINOS/MINOS+ experiment, which is important to constraint $\xi_{33}$ and consequently $|\xi_{31}|$ and $|\xi_{32}|$.  In our analysis, we found a better limit on the non-unitary parameter $|\xi_{21}|$, $\xi_{11}$ and $\xi_{22}$ compared to references \cite{Giunti_2021,Ellis_Kelly:2020}, in the ND developed case. Other bounds for all non-unitary parameters are presented in references \cite{Agarwalla:2021,Argawalla_2023}, where the latter was focused in constraints on $\xi_{32}$. Note that for all parameters that are constrained without the help of Cauchy-Schwarz inequalities (for diagonal parameters and $|\xi_{21}|$), the limits presented in our paper are very similar to the limits in ref. \cite{Agarwalla:2021}. However, $|\xi_{31}|$ and $|\xi_{32}|$ in our analysis have much stronger limits, also demonstrating how Cauchy-Schwarz inequalities plays a fundamental role.

\renewcommand\arraystretch{1.5}
\renewcommand\tabcolsep{0.4cm}
\begin{table}[ht]
\caption{Bounds at 90\% C.L. (1 d.o.f) on the non-unitary parameters obtained in this analysis from T2HK, T2HKK and DUNE when $N^{\mathrm{true}}$ is computed assuming unitary. The limits in the first line from each experiment was obtained under the ND undeveloped case and the second line under the ND developed case.}
\centering
\begin{tabular}{c c c c c c c}
\hline
\hline
 & $|\xi_{21}|$ & $|\xi_{31}|$& $|\xi_{32}|$ & 1 - $\xi_{11}^{2}$ & 1 - $\xi_{22}^{2}$ & 1 - $\xi_{33}^{2}$\\
\hline
\hline
T2HK & 0.016 & 0.15  & 0.058 & 0.13 & 0.018 & 0.45 \\
     & 0.010 & 0.097 & 0.057 & 0.047 & 0.016  & 0.45     \\
\hline
T2HKK & 0.014 & 0.11 & 0.052 & 0.094 & 0.014 & 0.43 \\ 
      & 0.0098 & 0.086 & 0.050 & 0.042 & 0.013 & 0.43 \\
\hline
DUNE & 0.015  & 0.078 & 0.052 & 0.051 & 0.025 & 0.21 \\
     & 0.0035 & 0.048 & 0.022 & 0.028 & 0.0067 & 0.21    \\
\hline
\hline
\end{tabular}
\label{tab:bounds}
\end{table}

\subsection{Case where $N^{\mathrm{true}}$ is computed assuming non-unitary}
\begin{figure}[ht]
\centering
\includegraphics[scale = 1.08]{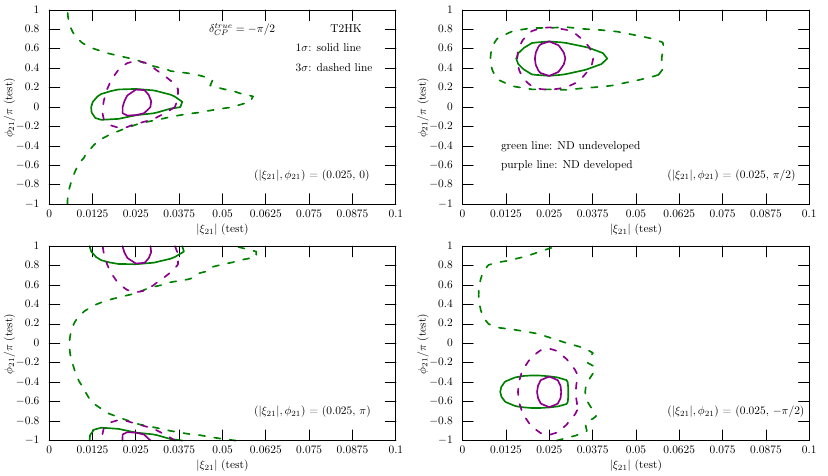}
\caption{1$\sigma$ and 3$\sigma$ (2 d.of.) C.L. contours in the ($|\xi_{21}|, \phi_{21}$) plane for T2HK, assuming ND undeveloped scenario (green line) and ND developed scenario (purple line). We have fixed $|\xi_{21}|^{\text{true}}$ = 0.025 and four values for $\phi_{21}^{\mathrm{true}}$. The values for other non-unitary parameters are described in Section \ref{sec:simu}.}
\label{fig:contourHK}
\end{figure}

\begin{figure}[ht]
\centering
\includegraphics[scale = 1.08]{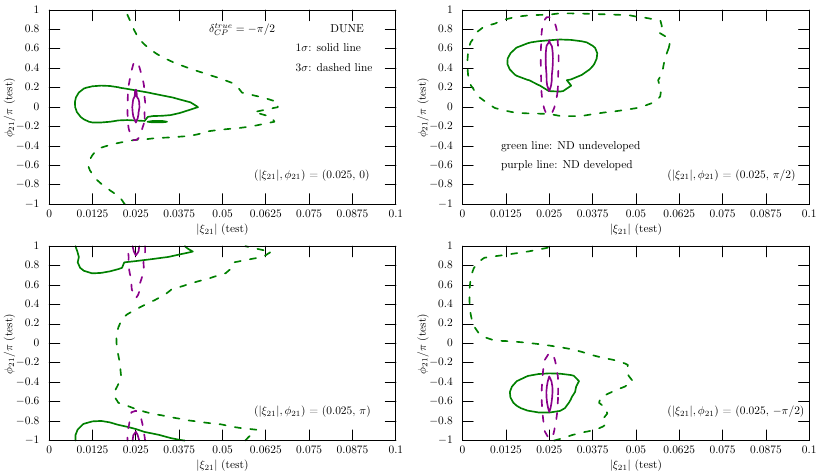}
\caption{1$\sigma$ and 3$\sigma$ (2 d.of.) C.L. contours in the ($|\xi_{21}|, \phi_{21}$) plane for DUNE, assuming ND undeveloped scenario (green line) and ND developed scenario (purple line). We have fixed $|\xi_{21}|^{\text{true}}$ = 0.025 and four values for $\phi_{21}^{\mathrm{true}}$. The values for other non-unitary parameters are described in Section \ref{sec:simu}.}
\label{fig:contourDUNE}
\end{figure}

Finally, we show the allowed regions for the $\xi_{21}$-$\phi_{21}$ plane. In order to do this, let us assume $|\xi_{21}|^{\mathrm{true}} = 0.025$, which lies within the current 3$\sigma$ allowed region, and the corresponding non-unitary phase, $\phi_{21}$, where we will consider four benchmarks $\phi_{21}^{\text{true}} = 0, \pi/2, \pi, -\pi/2$. All other non-unitarity parameters have been kept fixed at their unitary values, except those related to $\xi_{21}$ by Cauchy-Schwarz. The values used for these parameters were described in Section \ref{sec:simu}. The results of our analysis are shown in Figures \ref{fig:contourHK} and \ref{fig:contourDUNE}, for T2HK and DUNE, respectively. The bounds obtained by T2HKK are very similar to the results for T2HK and will not be shown. The green (purple) lines represent bounds obtained in the ND undeveloped (developed) scenario. 

From Figures \ref{fig:contourHK} and \ref{fig:contourDUNE}, the bounds on $|\xi_{21}|$ are more restrictive in the ND developed scenario. This is attributable to the fact that when non-unitarity exhibits short-range behavior, the near detector plays a crucial role in the sensitivity of this parameter. It is also worth noting that the limits derived for $\phi_{21}$ stem from the far detector, given that the near detector is insensitive to this particular phase. Tables \ref{tab:1sUncHK} and \ref{tab:1sUncDUNE} summarize the uncertainty on $|\xi_{21}|$ and $\phi_{21}$ for T2HK and DUNE, respectively.  

\begin{table}[ht]
\centering
\caption{Expected 1$\sigma$ (2 d.o.f.) uncertainty in the ($|\xi_{21}|, \phi_{21}$) plane for the ND undeveloped, the ND developed and using only the ND for T2HK. The true number of events was computed assuming non-unitarity with $|\xi_{21}|^{\mathrm{true}} = 0.025$ and four values for $\phi_{21}^{\mathrm{true}}$.}
\begin{tabular}{cccc}
\hline
\hline
 ($|\xi_{21}|^{\mathrm{true}}, \phi_{21}^{\mathrm{true}}$) & ND undeveloped & ND developed & only ND \\
 \hline
 \hline
 (0.025, 0) & (50\%, 23$^\circ$) & (16\%, 23$^\circ$) & (17\%, 180$^\circ$)\\
 \hline
 (0.025, 90$^\circ$) & (52\%, 32$^\circ$)  & (17\%, 32$^\circ$) & (17\%, 180$^\circ$)\\
 \hline
(0.025, 180$^\circ$) & (52\%, 23$^\circ$)  & (14\%, 23$^\circ$) & (17\%, 180$^\circ$) \\
\hline
(0.025, -90$^\circ$) & (39\%, 28$^\circ$) & (14\%, 28$^\circ$) & (17\%, 180$^\circ$)\\
\hline
\hline
\end{tabular}
\label{tab:1sUncHK}
\end{table}

\begin{table}[ht]
\centering
\caption{Expected 1$\sigma$ (2 d.o.f.) uncertainty in the ($|\xi_{21}|, \phi_{21}$) plane for the ND undeveloped, the ND developed and  using only the ND for DUNE. The true number of events was computed assuming non-unitarity with $|\xi_{21}|^{\mathrm{true}} = 0.025$ and four values for $\phi_{21}^{\mathrm{true}}$.}
\begin{tabular}{cccc}
\hline
\hline
 ($|\xi_{21}|^{\mathrm{true}}, \phi_{21}^{\mathrm{true}}$) & ND undeveloped & ND developed & only ND \\
 \hline
 \hline
 (0.025, 0) & (71\%, 28$^\circ$) &  (4.2\%, 28$^\circ$) & (4.2\%, 180$^\circ$)\\
 \hline
 (0.025, 90$^\circ$) & (51\%, 47$^\circ$)  & (4.2\%, 47$^\circ$) & (4.2\%, 180$^\circ$)\\
 \hline
(0.025, 180$^\circ$) & (62\%, 25$^\circ$)  & (3.8\%, 18$^\circ$) & (4.2\%, 180$^\circ$) \\
\hline
(0.025, -90$^\circ$) & (38\%, 36$^\circ$) & (3.5\%, 34$^\circ$) & (4.2\%, 180$^\circ$)\\
\hline
\hline
\end{tabular}
\label{tab:1sUncDUNE}
\end{table}

\section{CP Violation Sensitivity}\label{sec:CPSens}
One important goal of long-baseline neutrino experiments is to determine the value of the CP-violation phase, $\delta_{CP}$. In this work, our main focus is to analyze how non-unitarity mixing affects the determination of $\delta_{CP}$ at T2HK, T2HKK and DUNE considering the cases described in Section \ref{sec:simu}. Initially, the effects of unitarity violation and cross section uncertainty are analyzed individually. Subsequently, to enhance the realism of the scenario, these factors are examined in conjunction with one another. The results in this section were generated using the $\chi^{2}$ function  below, equation (\ref{eq:chi_squareCP}), where $\delta_{CP}^{\mathrm{true}}$ is varied in the range $[-\pi, \pi]$,
\begin{equation}
\chi_{\text{CPV}}^{2} =\underset{\vec{\eta}}{\mathrm{min}}  \left( \sum_{D}\sum_{k}^{\text{bins}} \frac{\left[N_{D, k}^{\text{true}}(\delta_{CP}^{\mathrm{true}}; \xi_{ij}) -N_{D, k}^{\text{fit}}(\delta_{CP}^{\mathrm{fit}} = 0 \hspace{1mm} \mathrm{or} \hspace{1mm}\pi; \xi_{ij}; \vec{\eta})\right]^{2}}{N_{D, k}^{\text{true}} (\delta_{CP}^{\mathrm{true}}; \xi_{ij})} + \sum_{n}\frac{\eta^{2}_{n}}{\sigma_{n}^{2}} \right).
\label{eq:chi_squareCP}
\end{equation} 

\begin{figure}[ht]
   \centering
    \includegraphics[scale=1.05]{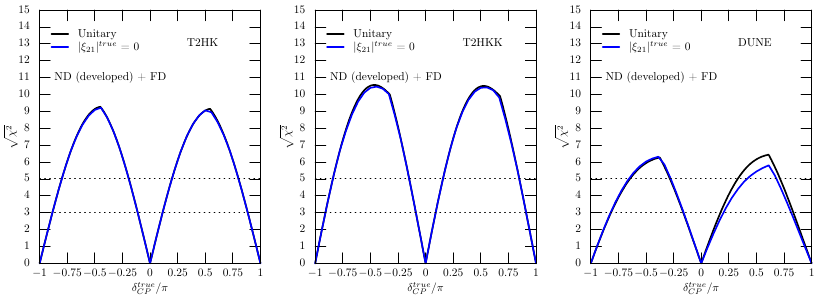}
    \caption{The ability of T2HK, T2HKK and DUNE to exclude CP conservation as a function of the true value of $\delta_{CP}$ is shown, respectively, by the left, middle and right panel for the ND developed case. Here $N^{\mathrm{true}}$ is computed assuming unitarity, and oscillation effects with unitarity violation have been averaged out both at the near and the far detector. The cross section uncertainty was not taken into account.}
    \label{fig:cpSens_NDFD-1a}
\end{figure}

\begin{figure}[ht]
   \centering
    \includegraphics[scale=1.05]{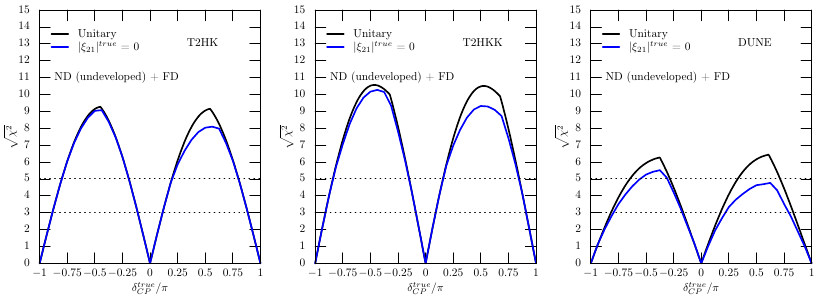}
    \caption{The ability of T2HK, T2HKK and DUNE to exclude CP conservation as a function of the true value of $\delta_{CP}$is shown, respectively, by the left, middle and right panel for the ND undeveloped case. Here $N^{\mathrm{true}}$ is computed assuming unitarity, and unitarity violation is averaged out only at the far detector. The cross section uncertainty was not taken into account.}
    \label{fig:cpSens_NDFD-2a}
\end{figure}

\begin{figure}[ht]
   \centering
    \includegraphics[scale=1.15]{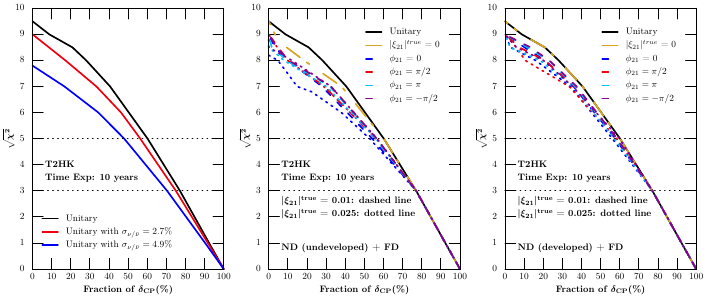}
    \caption{Signiﬁcance which the CP violation could be established by T2HK as a function of the fraction of values of $\delta_{CP}$ for which it would be possible. The horizontal dotted black line corresponds to 3$\sigma$ and 5$\sigma$. The left panel show the impact of the cross section uncertainty in the unitary scenario considering the current uncertainty (4.9\%) and the expected uncertainty (2.7\%). The middle (right) panel represent the impact of non-unitary in the ND undeveloped (ND developed) scenario. }
    \label{fig:cpFraction-HK}
\end{figure}

\begin{figure}[ht]
   \centering
    \includegraphics[scale=1.15]{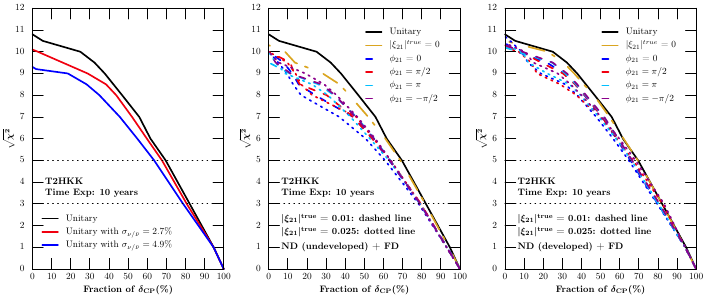}
    \caption{Signiﬁcance which the CP violation could be established by T2HKK as a function of the fraction of values of $\delta_{CP}$ for which it would be possible. The horizontal dotted black line corresponds to 3$\sigma$ and 5$\sigma$. The left panel show the impact of the cross section uncertainty in the unitary scenario considering the current uncertainty (4.9\%) and the expected uncertainty (2.7\%). The middle (right) panel represent the impact of non-unitary in the ND undeveloped (ND developed) scenario.}
    \label{fig:cpFraction-HKK}
\end{figure}

\begin{figure}[ht]
   \centering
    \includegraphics[scale=1.15]{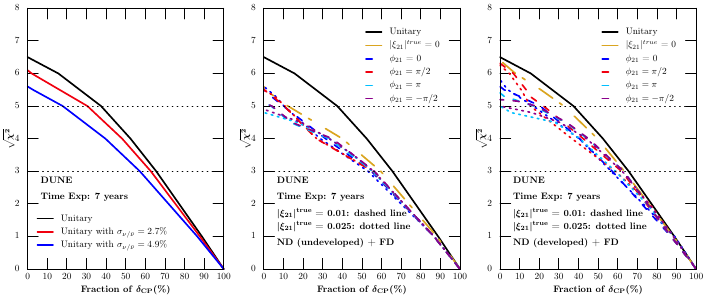}
    \caption{Signiﬁcance which CP violation could be established by DUNE as a function of the fraction of values of $\delta_{CP}$ for which it would be possible. The horizontal dotted black line corresponds to 3$\sigma$ and 5$\sigma$. The left panel show the impact of the cross section uncertainty in the unitary scenario considering the current uncertainty (4.9\%) and the expected uncertainty (2.7\%). The middle (right) panel represent the impact of non-unitary in the ND undeveloped (ND developed) scenario.}
    \label{fig:cpFraction-DUNE}
\end{figure}

Already anticipating some comments that will be made below, we did not find considerable differences between all the proposed situations for the non-unitarity scenarios, therefore, we will only show in Figures \ref{fig:cpSens_NDFD-1a} and \ref{fig:cpSens_NDFD-2a}, the situation (a) $N^{\mathrm{true}}$ is computed assuming unitarity, for the two cases ND developed and ND undeveloped and for all the experiments considered. Note that this does not necessarily mean that we will be assuming that there is no non-unitarity effect at all but we are also covering the case where there can be some subdominant effects of non-unitarity but they are small enough such that the experimental data would be considered consistent with the unitarity conservation. In order to summarize the results of the CP sensitivity, Figures \ref{fig:cpFraction-HK}, \ref{fig:cpFraction-HKK} and \ref{fig:cpFraction-DUNE} show the fraction of values of $\delta_{CP}$ for which a given signiﬁcance for CP could be established considering the impact of the cross section uncertainty and the impact of all the proposed cases for violation of unitarity, for T2HK, T2HKK and DUNE, respectively. We show in the left panel of each figure the unitary case over the influence of cross section uncertainty considering the current uncertainty 4.9\% and the expected uncertainty 2.7\%. The middle (right) panel is considering unitarity violation for the ND undeveloped scenario (ND developed). Some comments are worth making:
\begin{itemize}
\item For T2HK and T2HKK, the reduction of CP sensitivity due to unitarity violation is very small, and the experiments can still establish CP violation above 5$\sigma$ for the two considered cases (ND developed and undeveloped). In contrast, DUNE is the most influenced by non-unitary parameters. The ability to measure the CP sensitivity changes mainly in the ND undeveloped case. In the ND developed case, the combination of ND + FD can eliminate the dependence of the probability on some non-unitary parameters, which may explain the lack of reduction in the CP sensitivity. It is important to emphasize that the reduction observed in Figure \ref{fig:cpSens_NDFD-2a} for DUNE does not stem from the matter effects. We have verified that if DUNE were in a vacuum, this reduction would still occur.
\item For the three situations proposed in Section \ref{sec:simu} in the unitarity violation scenario, we found no significant discrepancies between cases (a) no unitariy violation in input, (b) intermediate case and (c) upper limit case, either for the ND developed or for the ND undeveloped scenario. In other words, the impact of the three cases is very similar, and within the systematic uncertainties we are unable to separate them, even when the value of the non-unitary parameter $|\xi_{21}|$ is considered large (0.025). 
\item  In the unitary scenario, T2HK and DUNE are more susceptible to the effects of cross section uncertainty than T2HKK. This result appears to align with the results presented in ref. \cite{Ghosh-2017}, in the standard oscillation scenario adopted by the paper. Comparing the sensitivity reduction of the unitary case (with $\sigma_{\nu/\bar{\nu}}$= 4.9\%) with the non-unitary cases, the following observations can be made: the uncertainty in the cross section is significantly more detrimental than non-unitarity in T2HK and T2HKK. On the other hand, DUNE suffers more from the impact of non-unitarity.
\end{itemize}


\begin{figure}[ht]
   \centering
    \includegraphics[scale=1.05]{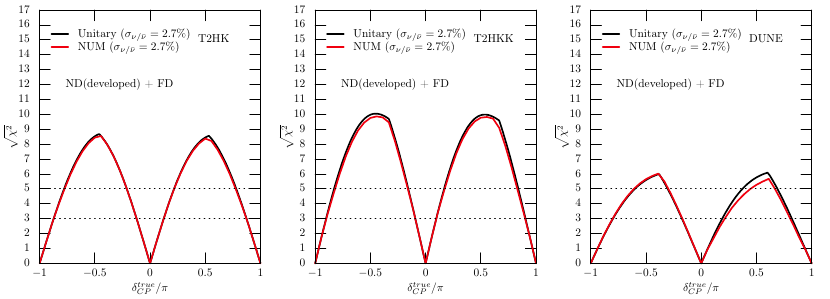}
    \caption{The ability of T2HK, T2HKK and DUNE to exclude CP conservation as a function of the true value of $\delta_{CP}$ in the ND developed scenario. Here $N^{\mathrm{true}}$ is computed assuming unitarity. The black line is the unitary case with the expected (2.7\%) cross section uncertainty. The red lines represents the impact of unitarity violation and cross section uncertainty (2.7\%).}
    \label{fig:cpSens_NDFD-1a-NUM_XSEC}
\end{figure}

\begin{figure}[ht]
   \centering
    \includegraphics[scale=1.05]{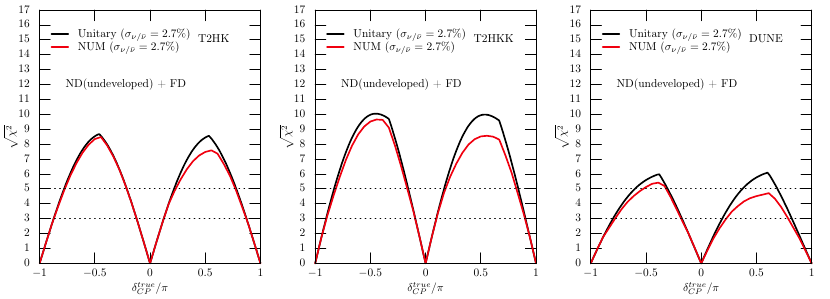}
    \caption{The ability of T2HK, T2HKK and DUNE to exclude CP conservation as a function of the true value of $\delta_{CP}$ in the ND undeveloped scenario. Here $N^{\mathrm{true}}$ is computed assuming unitarity. The black line is the unitary case with the expected (2.7\%) cross section uncertainty. The red lines represents the impact of unitarity violation and cross section uncertainty (2.7\%).}
    \label{fig:cpSens_NDFD-2a-NUM_XSEC}
\end{figure}

In order to achieve a more accurate representation of reality, unitarity violation and cross section uncertainty must be considered together. The results are presented in Figure \ref{fig:cpSens_NDFD-1a-NUM_XSEC} for the ND developed scenario and Figure \ref{fig:cpSens_NDFD-2a-NUM_XSEC} for the ND undeveloped scenario, where $N^{\mathrm{true}}$ is computed assuming unitarity conservation. The conclusions drawn here could already be anticipated based on previous results. However, to summarize, for T2HK and T2HKK, the main contribution to reduction of sensitivity is due to cross section uncertainty, these experiments are robust over non-unitarity impact. For DUNE, the main cause of reduction in CP sensitivity is non-unitarity.  


\begin{figure}[ht]
  \centering
    \includegraphics[scale=1.15]{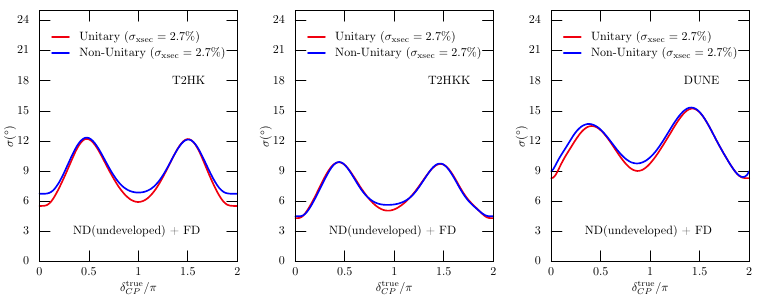}
    \caption{The 1$\sigma$ uncertainty in the allowed $\delta_{CP}$ range for T2HK (left), T2HKK (middle) and DUNE (right) as function of true $\delta_{CP}$. The red lines correspond to the unitary case with a cross section uncertainty of 2.7\%. The blue line represents the impact of non-unitarity in the ND undeveloped scenario and cross section with an uncertainty of 2.7\%.}
    \label{fig:precision}
\end{figure}

Finally, the impact of non-unitarity on $\delta_{CP}$ measurement precision in 1$\sigma$ ($\sqrt{\chi^{2}} = 1$) is shown in Figure \ref{fig:precision}. We have assumed the ND undeveloped case for non-unitarity and $N^{\mathrm{true}}$ is computed assuming unitarity, this scenario is represented by the blue lines in this figure. The red lines represent the unitary case with a cross section uncertainty of 2.7\%. We can see that T2HKK has the best precision to measure $\delta_{CP}$, followed by T2HK and DUNE \footnote{The results found here for the standard case differ quantitatively from those officially presented by the collaborations \cite{DUNE,T2HK,T2HKK}. The reason is that here we are doing a simplified treatment in relation to the experiment and its uncertainties, thus reflecting in the results found. However, qualitatively, the results are broadly consistent.}. The sensitivity is better at $\delta_{CP} = 0, \pm 180^{\circ}$ and worst close to $\delta_{CP} = \pm 90^{\circ}$. In the standard case, T2HK has an uncertainty of $6^{\circ} - 13^{\circ}$ for most values of $\delta_{CP}$, for T2HKK the uncertainty is around $4^{\circ} - 10^{\circ}$ and finally for DUNE, resolutions between $8^{\circ} - 14^{\circ}$ are possible, depending on the true value of $\delta_{CP}$. As we can see in Figure \ref{fig:precision}, non-unitarity does not have any significant impact on the measurements of $\delta_{CP}$, thus showing how robust the experiments are.

\section{Conclusion}\label{sec:conclusion}

As we are entering in the higher precision era of neutrino experiments, we have a good chance of testing the unitarity of the leptonic mixing matrix. Furthermore, if non-unitarity exists, it is important to verify how this scenario affects the sensitivity of the standard oscillation parameters. In this work, we have used explicit near and far detector simulations, which have not been done in most of the previous studies, and we assumed two scenarios for unitarity violation. The first one is when the unitarity violation manifests itself both at the near and far detectors, called the ND developed scenario. In this scenario, the near detector can play a fundamental role in the sensitivity of the new parameters as it becomes sensitive to new physics. The second one is called the ND undeveloped scenario, where the unitarity violation has not yet manifested at near detector, only manifested at the far detector, thus, the near detector measures the standard flux and cross section. Taking into account these two approaches, our main goal with this work is to assess the impact of each situation on the sensitivity of non-unitarity parameters and the standard CP phase, $\delta_{CP}$, for T2HK, T2HKK and DUNE, the next generation of long-baseline neutrino experiment.

The sensitivity of the non-unitarity parameters obtained in this work, under the two aforementioned scenarios, is presented in Figure \ref{fig:projection}. The bounds in the ND developed scenario are better compared to the ND undeveloped case. These results differ from those presented in ref. \cite{Blennow_Pila:2016}. The previous studies normalized the probability equation to account for the near detector effect,  when this is done, the primary conclusion is that the sensitivity of some non-unitary parameters is reduced in the ND developed case; our results demonstrate that this reduction does not occur. For the $\xi_{11}$ parameter, constrained mainly by the appearance channel, the influence of the background channel $\nu_{e} \rightarrow \nu_{e}$ is extremely important for the sensitivity. Furthermore, we are taking into account an important mathematical condition, Cauchy-Schwarz inequalities, condition that guarantees the constraint of the off-diagonal non-unitary parameters, even if the experiment is not so sensitive to them. Therefore, an increase in the sensitivity of the diagonal parameter consequently influences a better sensitivity in the off-diagonal parameter related by this mathematical condition.

Regarding the sensitivity of non-unitary parameters, Figures \ref{fig:contourHK} and \ref{fig:contourDUNE} display the sensitivity when $|\xi_{21}|$ is non-zero and the true number of events is computed assuming non-unitary scenario. The results highlight a notable improvement in the sensitivity of $|\xi_{21}|$ when the ND developed scenario is considered. The primary contributor to the sensitivity of $|\xi_{21}|$ is the near detector, while the non-unitary phase $\phi_{21}$ is constrained solely from the far detector.

In order to estimate the impact of unitarity violation on the sensitivity of $\delta_{CP}$, we considered three scenarios. Firstly, the case where $N^{\mathrm{true}}$ is computed assuming unitarity, secondly, $N^{\mathrm{true}}$ is computed assuming non-unitarity with $|\xi_{21}|^{\mathrm{true}}$ = 0.01, and thirdly, $N^{\mathrm{true}}$ is computed assuming non-unitarity but with $|\xi_{21}|^{\mathrm{true}}$ = 0.025 and $\phi_{21}$ assuming four benchmarks $[0, \pi/2, \pi, -\pi/2]$ in the latter two scenarios. In order to get closer to reality, the uncertainty in the cross section was also taken into account, such systematic is the dominant source of uncertainty in the determination of $\delta_{CP}$ phase, mainly to T2HK. The impact of non-unitarity on the sensitivity test to exclude $\sin \delta_{CP} = 0$, when $N^{\mathrm{true}}$ is computed assuming unitarity can be observed in Figures \ref{fig:cpSens_NDFD-1a} and \ref{fig:cpSens_NDFD-2a}, for the ND developed and the ND undeveloped case, respectively. We did not observe any truly significant differences among the three cases considered in terms of their impact on sensitivity, the summary of the results is shown in Figures \ref{fig:cpFraction-HK}, \ref{fig:cpFraction-HKK}
 and \ref{fig:cpFraction-DUNE}, each of them showing the significance that the CP violation could be established by T2HK, T2HKK and DUNE, respectively, as a function of the fraction of values of $\delta_{CP}$ for which it would be possible. T2HK and T2HKK are robust against the impact of unitarity violation, reaching 5$\sigma$ in all scenarios considered here. For these experiments the uncertainty in the cross-section is the most significant factor reducing the sensitivity, particularly for T2HK. On the other hand, for DUNE, non-unitarity plays a primary role in reducing the sensitivity, however, such reduction depends on the non-unitary scenario considered. In the ND developed case, the sensitivity is almost completely recovered, reaching 5$\sigma$ when $N^{\mathrm{true}}$ is computed assuming unitarity. 

\appendix
\section{Event Rates}\label{AppendixA}
\begin{figure}[ht]
\centering
\includegraphics[scale = 1.15]{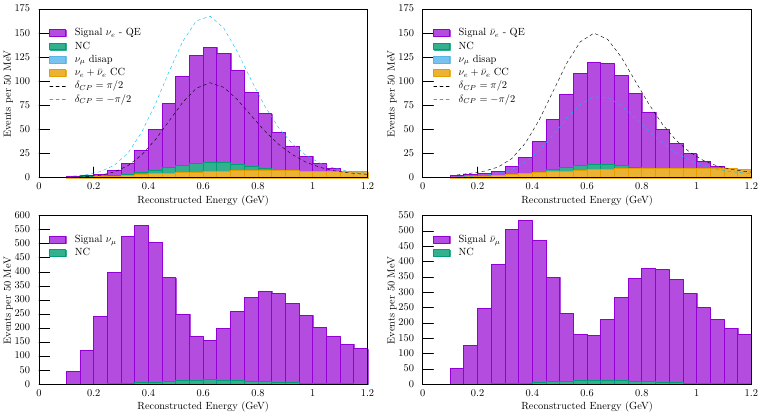}
\caption{Reconstructed neutrino energy distribution of the $\nu/\bar{\nu}$ events for T2HK in the neutrino (left) and antineutrino (right) mode, assuming unitarity. The neutrino signal is represented by the purple filled region, with $\delta_{CP} = 0$. The blue and black lines are the neutrino signal but for $\delta_{CP} = \pi/2$ and $\delta_{CP} = -\pi/2$, respectively. The background is represented by colored regions.}
\label{fig:events_hk}
\end{figure}

\begin{figure}[ht]
\centering
\includegraphics[scale = 1.15]{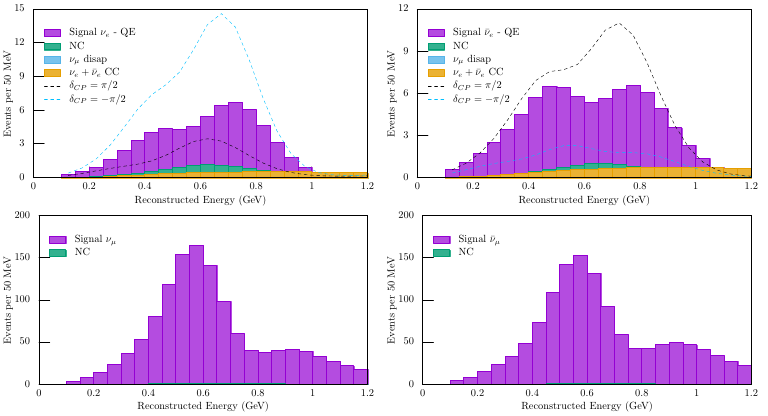}
\caption{Reconstructed neutrino energy distribution of the $\nu/\bar{\nu}$ events for KD in the neutrino (left) and antineutrino (right) mode, assuming unitarity. The neutrino signal is represented by the purple filled region, with $\delta_{CP} = 0$. The blue and black lines are the neutrino signal but for $\delta_{CP} = \pi/2$ and $\delta_{CP} = -\pi/2$, respectively. The background is represented by colored regions.}
\label{fig:events_kd}
\end{figure}

\begin{figure}[ht]
\centering
\includegraphics[scale = 1.15]{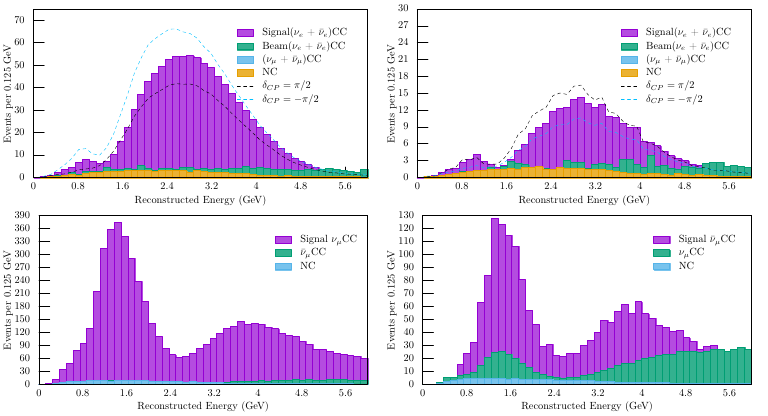}
\caption{Reconstructed neutrino energy distribution of the $\nu/\bar{\nu}$ events for DUNE in the neutrino (left) and antineutrino (right) mode, assuming unitarity. The neutrino signal is represented by the purple filled region, with $\delta_{CP} = 0$. The blue and black lines are the neutrino signal but for $\delta_{CP} = \pi/2$ and $\delta_{CP} = -\pi/2$, respectively. The background is represented by colored regions.}
\label{fig:events_dune}
\end{figure}

In this section, we present the number of events considered during our simulation and analysis. For T2HK/T2HKK, we are following ref. \cite{T2HK,T2HKK}. For DUNE, we are using the files provided by the collaboration, ref. \cite{DUNEGlobes}. The simulated event rates are shown in Figures \ref{fig:events_hk}, \ref{fig:events_kd} and \ref{fig:events_dune}, for T2HK, KD (only detector in Korea) and DUNE, respectively. The figures show only results for the unitary mode.  The oscillation parameters used are shown in Table \ref{tab:std_osc}. 

The background for T2HK/KD are the beam contamination, the misidentified $\nu_{\mu}$ and the neutral current (NC) in both the appearance and disappearance channels. For DUNE, the background of the appearance channel are the beam contamination, the misidentified $\nu_{\mu}$ and the NC. In the disappearance channel both, the NC and the misidentified $\nu_{\mu}$ compose the background.

For the near detector, the number of events in the unitary mode, is of the order of $\mathcal{O}(10^{7})$ for T2HK/T2HKK and 
$\mathcal{O}(10^{8})$ for DUNE,  in the $\nu_{\mu}$ ($\nu_{\bar{\mu}}$) channel. In the non-unitary regime, electron neutrino appearance events are possible at short distances. Thus, for T2HK/T2HKK we expect $\mathcal{O}(10^{4})$ $\nu_{e}$ and $\nu_{\bar{e}}$ events. For DUNE, in the $\nu_{e}$-channel we expect $\mathcal{O}(10^{5})$ events and $\mathcal{O}(10^{4})$ for $\nu_{\bar{e}}$ events.
\section{Exact Equations of Neutrino Oscillations in Vacuum}\label{AppendixB}

In this appendix, we write the exact equations for muon neutrino appearance and disappearance in vacuum, considering the scenario of three neutrinos flavors. We have used the standard parametrization of the PMNS mixing matrix \cite{ParticleDataGroup:2024} and defined $\Delta_{kj} = \Delta m_{kj}^{2}L/2E$ with $\Delta m_{kj}^{2} = m_{k}^{2} - m_{j}^{2}$, $L$ being the distance between the source and the detector and $E$ being the neutrino energy. The appearance equation is given by,
\begin{align}
&\quad P_{\mu e}^{3 \times 3} = -4\Bigg\{-\cos\theta_{12}\sin\theta_{23}\sin\theta_{13}\cos^{2}\theta_{13}(\cos\theta_{12}\sin\theta_{23}\sin\theta_{13} \nonumber \\ 
&\quad + \sin\theta_{12}\cos\theta_{23}\cos\delta_{CP})\sin^{2}\left(\frac{\Delta_{31}}{2}\right) - \sin\theta_{12}\sin\theta_{23}\sin\theta_{13}\cos^{2}\theta_{13}\nonumber \\
&\quad \times (\sin\theta_{12}\sin\theta_{23}\sin\theta_{13} - \sin\theta_{12}\cos\theta_{23}\cos\delta_{CP})\sin^{2}\left(\frac{\Delta_{32}}{2}\right) \nonumber \\
&\quad - \frac{1}{4}\sin(2\theta_{12})\cos^{2}\theta_{13}\Big[(\cos^{2}\theta_{23} - \sin^{2}\theta_{23}\sin^{2}\theta_{13})\sin(2\theta_{12}) \nonumber \\
&\quad + \sin(2\theta_{23})\cos(2\theta_{12})\sin\theta_{13}\cos\delta_{CP}\Big] \sin^{2}\left(\frac{\Delta_{21}}{2}\right) \Bigg\}  \nonumber \\
&\quad + 2\cos\theta_{13}\Bigg\{ - \sin\theta_{12}\cos\theta_{12}\sin\theta_{23}\cos\theta_{23}\sin\theta_{13}\cos\theta_{13}\sin\delta_{CP}\sin\Delta_{31} \nonumber \\
&\quad + \sin\theta_{12}\cos\theta_{12}\sin\theta_{23}\cos\theta_{23}\sin\theta_{13}\cos\theta_{13}\sin\delta_{CP}\sin\Delta_{32} \nonumber \\
&\quad + \sin\theta_{12}\cos\theta_{12}\sin\theta_{23}\cos\theta_{23}\sin\theta_{13}\cos\theta_{13}\sin\delta_{CP}\sin\Delta_{21} \Bigg\}.
\end{align}

The disappearance equation is given by,
\begin{align}
&\quad P_{\mu \mu}^{3 \times 3} = 1 
- 4 \Bigg\{\Bigg[\sin^{2}\theta_{23}\cos\theta_{13} \Big(\sin^{2}\theta_{12}\cos^{2}\theta_{23} +
\cos^{2}\theta_{12}\sin^{2}\theta_{23}\sin^{2}\theta_{13} +
\nonumber \\
&\quad + \frac{1}{2}\sin(2\theta_{12})\sin(2\theta_{23})\sin\theta_{13}\cos\delta_{CP}\Big)\Bigg] \sin^{2}\left(\frac{\Delta_{31}}{2}\right)+ \nonumber \\
&\quad + \Bigg[ \sin^{2}\theta_{23}\cos^{2}\theta_{13} \Bigg(\cos^{2}\theta_{12}\cos^{2}\theta_{23} + \sin^{2}\theta_{12}\sin^{2}\theta_{23}\sin^{2}\theta_{13}- \nonumber \\
&\quad - \frac{1}{2}\sin(2\theta_{12})\sin(2\theta_{23})\sin\theta_{13}\cos\delta_{CP}\Bigg) \Bigg] 
\sin^{2}\left(\frac{\Delta_{32}}{2}\right)+ \nonumber \\
&\quad + \Bigg[ \Bigg( 1 - \frac{1}{2}\sin^{2}(2\theta_{12})\Bigg)\frac{1}{4}\sin^{2}(2\theta_{23})\sin^{2}\theta_{13}  - \frac{1}{4}\sin^{2}(2\theta_{12})\sin^{2}(2\theta_{23})\sin^{2}\theta_{13}\cos^{2}\delta_{CP}+   \nonumber \\
&\quad + \frac{1}{4}\sin(4\theta_{12})\sin(2\theta_{23})(\cos^{2}\theta_{23} - \sin^{2}\theta_{23}\sin^{2}\theta_{13})\sin\theta_{13}\cos\delta_{CP}+ \nonumber \\
&\quad + \frac{1}{4} \sin^{2}(2\theta_{12})(\cos^{4}\theta_{23} + \sin^{4}\theta_{23}\sin^{4}\theta_{13})\Bigg]\sin^{2}\left(\frac{\Delta_{21}}{2}\right)\Bigg\}.
\end{align}

\acknowledgments

This study was financed in part by the Coordenação de Aperfeiçoamento de Pessoal de Nível Superior - Brasil (CAPES) - Finance Code 001 (scholarship No. 88887.479307/2020-00 and grant No. 312263/2021-0) and Conselho Nacional de Desenvolvimento Científico e Tecnológico (CNPq) scholarship No. 166794/2022-8.



\bibliographystyle{JHEP}
\bibliography{biblio.bib}






\end{document}